\documentclass{aa}
\usepackage{graphicx}
\usepackage{txfonts}
\begin{document}
\setlength{\textheight}{684pt} 
\newdimen\digitwidth
\newcommand {\magg}{\hphantom{$>$}}  
\setbox0=\hbox{\rm0}
\digitwidth=\wd0
\catcode `#=\active
\def#{\kern\digitwidth}

   \titlerunning{ISO Observations of 1 Jy BL Lacertae Objects} 

   \title{Filling the Infrared Gap: \\ISO Observations of 1 Jy BL Lacertae 
Objects}


   \author{Paolo Padovani
          \inst{1}
          \and
          Paolo Giommi\inst{2}
          \and
           P\'eter \'Abrah\'am\inst{3}
           \and
           Szil\'ard Csizmadia\inst{3}
           \and
           Attila Mo\'or\inst{3}     
          }

   \offprints{Paolo Padovani}

   \institute{European Southern Observatory, Karl-Schwarzschild-Str. 2,
             D-85748 Garching bei M\"unchen, Germany\\
              \email{Paolo.Padovani@eso.org}
         \and
             ASI Science Data Center, ASDC, Agenzia Spaziale Italiana c/o 
             ESRIN, Via G. Galilei, I-00044 Frascati, Italy\\
             \email{paolo.giommi@asdc.asi.it}
         \and
              Konkoly Observatory, H-1525 Budapest, P.O. Box 67, Hungary\\
              \email{abraham@konkoly.hu, csizmadi@konkoly.hu, moor@konkoly.hu}
          }
   \date{Received ; accepted}

\abstract
{}
{The large majority of BL Lacertae objects belonging to the 1 Jy sample,
the class prototype for radio-selected sources, are thought to emit most of
their synchrotron power in the far infrared band. Ironically, this spectral
region is very sparsely sampled, with only a minority of the objects having
IRAS data (most of them being upper limits or low-quality detections). We
aim at filling this infrared gap by presenting new, simultaneous ISOCAM and
ISOPHOT observations over the $7 - 200\mu$m range (observer's frame) for
half the sample. A precise measurement of the position of the synchrotron
peak frequency, $\nu_{\rm peak}$, can provide direct information about
particle acceleration mechanisms and constrain the inverse Compton
radiation that will be detected by up-coming new $\gamma$-ray missions.}
{We have observed seventeen 1 Jy BL Lacertae objects with the camera and
the photometer on board the Infrared Space Observatory (ISO) satellite.
Given the intrinsic variability of these sources, the data were taken by
concatenating the pointings to ensure simultaneity. The ISOPHOT data
reduction was done employing a novel correction, which mitigates the effect
of chopping for faint sources.}
{Using our new ISO data, complemented by nearly-simultaneous radio and
optical observations for ten and four objects respectively, and other
multi-frequency data, we have built the spectral energy distributions of
our sources (plus a previously published one) and derived the rest-frame
$\nu_{\rm peak}$. Its distribution is centered at $\sim 10^{13}$ Hz ($\sim
30\mu$m) and is very narrow, with $\sim 60\%$ of the BL Lacs in the $1 - 3
\times 10^{13}$ Hz range. Given our set of simultaneous infrared data,
these represent the best determinations available of the synchrotron peak
frequencies for low-energy peaked BL Lacs. A comparison with previous such
estimates, based on non-simultaneous optical and near infrared data, may
indicate strong $\nu_{\rm peak}$ variations in a number of sources,
possibly associated with large flares as observed in the high-energy peaked
BL Lac MKN 501.}
{}
   

   \maketitle
%

\section{Introduction}

BL Lacertae objects constitute one of the most extreme classes of active
galactic nuclei (AGN), distinguished by their high luminosity, rapid
variability, high ($> 3\%$) optical polarization, radio core-dominance,
apparent superluminal speeds, and almost complete lack of emission lines
(e.g., Urry \& Padovani \cite{UP95}). The broad-band emission in these
objects, which extends from the radio to the gamma-ray band, appears to be
dominated by non-thermal processes from the heart of the AGN, undiluted by
the thermal emission present in other AGN. Therefore, BL Lacs represent the
ideal class to study to further our understanding of non-thermal emission
in AGN.

Synchrotron emission combined with inverse Compton scattering is generally
thought to be the mechanism responsible for the production of radiation
over such a wide energy range (e.g., Ghisellini et al.
\cite{Ghisellini98}). The frequency at which most of the synchrotron power
is emitted, $\nu_{\rm peak}$, ranges across several orders of magnitude,
going from the far-infrared to the hard X-ray band. Sources at the extremes
of this wide distribution are referred to as low-energy peaked (LBL) and
high-energy peaked (HBL) BL Lacs, respectively (Giommi \& Padovani
\cite{Giommi94}; Padovani \& Giommi \cite{Padovani95}). Radio-selected
samples include mostly objects of the LBL type, while X-ray selected
samples are mostly made up of HBL.

\begin{table*}
\caption{Sample properties.\label{prop}}
\begin{center}
\begin{tabular}{lrrllrl}
\hline
Name & RA(J2000) & Dec(J2000)&~~~$z$ &$\langle R_{\rm mag}\rangle$&
F$_{\rm 5GHz}$ \\
     &         &        &    &   &Jy \\
\hline
{PKS 0048$-$097} & 00 50 41.3 & $-$09 29 06 &$>$0.2 & 16.5 & 2.0 \\
{PKS 0118$-$272} & 01 20 31.7 & $-$27 01 24 &$>$0.559 & 17.0 & 1.1 \\
{PKS 0138$-$097} & 01 41 25.8 & $-$09 28 44 &\magg0.733 & 17.0 & 1.2 \\
{PKS 0426$-$380} & 04 28 40.4 & $-$37 56 19 &\magg1.110 & 18.0 & 1.2 \\
{S5 0454+844}    & 05 08 42.4 & $+$84 32 05 &$>$1.340 & 19.0 & 1.4 \\
{PKS 0537$-$441} & 05 38 50.4 & $-$44 05 09 &\magg0.896& 13.5 & 4.0 \\ 
{PKS 0735+178}   & 07 38 07.1 & $+$17 42 27 &$>$0.424 & 14.5 & 2.0 \\
{PKS 1144$-$379} & 11 47 01.4 & $-$38 12 10 &\magg1.048 & 16.5 & 1.6 \\
{OQ 530}         & 14 19 46.6 & $+$54 23 15 &\magg0.152 & 14.5 & 1.1 \\
{PKS 1519$-$273} & 15 22 37.7 & $-$27 30 10 &$>$0.2 & 18.5 & 2.4 \\
{PKS 1749+701}   & 17 48 32.8 & $+$70 05 51 &\magg0.770 & 16.5 & 1.1 \\
{PKS 1749+096}   & 17 51 32.8 & $+$09 39 01 &\magg0.320 & 15.5 & 1.9 \\ 
{S5 1803$+$784}  & 18 00 45.7 & $+$78 28 04 &\magg0.684 & 16.5 & 2.6 \\
{4C 56.27}       & 18 24 07.2 & $+$56 51 00 &\magg0.664 & 18.5 & 1.7 \\
{PKS 2131$-$021} & 21 34 10.3 & $-$01 53 17 &\magg1.285 & 20.0 & 2.1 \\
{PKS 2240$-$260} & 22 43 26.4 & $-$25 44 31 &\magg0.774 & 17.5 & 1.0 \\
{PKS 2254+074}   & 22 57 17.3 & $+$07 43 12 &\magg0.190 & 17.0 & 1.2 \\ 
\hline
\end{tabular}
\end{center}
\end{table*}

It is ironic that the spectrum of LBL in the infrared (IR) band, where
these sources' output is thought to be largest, is very sparsely known
(e.g., Sambruna et al. \cite{Sambruna96}; Giommi et al. \cite{Giommi02};
Massaro et al. \cite{Massaro05}). IR data for the 1 Jy, the prototype
radio-selected BL Lac sample are, for example, very sparse. The purpose of
this paper is to remedy this situation by presenting ISOCAM and ISOPHOT
observations of 1 Jy BL Lacs covering the $6.7 - 200\mu$m range. Since
$\nu_{\rm peak} \propto \gamma^2_{\rm peak} \delta B$, where $\gamma_{\rm
peak}$ is the Lorentz factor of the electrons emitting most of the
radiation, $\delta$ is the Doppler factor, and $B$ is the strength of the
magnetic field, its precise measurement can provide direct information
about particle acceleration mechanisms. The position of $\nu_{\rm peak}$ is
also important for the up-coming Astro-rivelatore Gamma a Immagini Leggero
(AGILE\footnote{\tt http://agile.rm.iasf.cnr.it/}) and Gamma ray Large Area
Space Telescope (GLAST\footnote{\tt http://glast.gsfc.nasa.gov/}) missions,
as they will sample the $\gamma$-ray band close to the peak of the inverse
Compton emission from LBL and flat-spectrum radio quasars, whose location
depends on that of the synchrotron one.

In Sect. 2 we present our sample, Sect. 3 discusses the Infrared Space
Observatory (ISO) observations, data analysis, and results, while in
Sect. 4 we derive the spectral energy distributions and $\nu_{\rm peak}$
values. Finally, Sect. 5 summarizes our conclusions. Throughout this paper
spectral indices are written $S_{\nu} \propto \nu^{-\alpha}$.

\section{The sample}

The 1 Jy sample of BL Lacs is the only sizeable, complete sample of radio
bright BL Lacs. It includes 34 objects with radio flux $> 1$ Jy at 5 GHz
(Stickel et al. \cite{Stickel91}). All 1 Jy BL Lacs have been studied in
detail in the radio and optical bands; all objects have also soft X-ray
data, primarily from ROSAT (Urry et al. \cite{Urry96}). Most of the sources
have also been detected by the Wilkinson Microwave Anisotropy Probe (WMAP)
satellite (Bennett et al. \cite{Bennett03}) and about $40\%$ of them have
also hard X-ray ({\it BeppoSAX}) data (e.g., Padovani et al.
\cite{Padovani04} and references therein). In the IR band, on the other
hand, 1 Jy BL Lacs are not very well studied. For example, only $\sim 40\%$
of 1 Jy BL Lacs have IRAS data, and many of these are of low quality or
only upper limits. We selected for ISO observations all 1 Jy BL Lacs
visible by the satellite and not already part of other programs. This
included seventeen 1 Jy BL Lacs (or $50\%$ of the sample). All sources are
LBL. The object list and basic characteristics are given in Tab.
\ref{prop}. Accurate radio positions and redshifts come from the NASA/IPAC
Extragalactic Database (NED), mean $R$ magnitudes are from Heidt \& Wagner
(\cite{Heidt96}), while 5 GHz radio fluxes are from Stickel et
al. (\cite{Stickel91}).

\section{Observations, data analysis, and results}

The ISO satellite (Kessler et al. \cite{Kessler96}) was equipped with a 60
cm Ritchey-Chr\'etien telescope and had four instruments. For our
observations we used the camera (ISOCAM) and the photometer (ISOPHOT).
Given the intrinsic variability of BL Lacs, we needed to observe each
object at a single epoch. Thus, the three different Astronomical
Observation Templates (AOTs) we used (see below) were concatenated in order
to warrant as much simultaneity as possible for the measurements.

\subsection{ISOCAM} 

All sources were observed with the $32 \times 32$ pixel ISOCAM (Blommaert
et al. \cite{Blommaert03}) in staring mode (AOT = CAM01) using the
long-wavelength (LW) detector in three different bands: LW2 ($\lambda =
6.7\mu$m), LW7 ($\lambda = 9.6\mu$m), and LW3 ($\lambda = 14.3\mu$m). The
pixel scale was $6^{\prime\prime}$ and the total field of view $196 \times
196$ arcsec$^2$. The total exposure times were computed as $T_{\rm int}
\times N_{\rm exp}$, where $T_{\rm exp} = 2.1$s is the integration time of
a single exposure and $N_{\rm exp}$ is the number of exposures. Typical
exposure times are 100\,s, 100\,s, and 75\,s for the LW2, LW7, and LW3
filters, respectively. A journal of the observations is presented in Tab.
\ref{isoobs}, which gives the name of the source, the corresponding ISO\_id
which uniquely identifies an ISO observation, and the observing date.

The data analysis was performed with the ISOCAM Interactive Analysis (CIA;
Ott et al. \cite{Ott97}) tool, Ver. 5.0. The default steps of CIA were used
for dark current subtraction, removal of cosmic ray hits (glitches),
exposure co-addition, flat fielding, and flux conversion to astronomical
units\footnote{This last step assumes $f_{\nu} \propto \nu^{-1}$, which is
the average spectrum of our sources (see below), so no colour-correction
was applied (in the most extreme case, $f_{\nu} \propto \nu^{-2.7}$, this
would have been $\sim 4\%$, way below our uncertainties).}. The transient
correction we adopted was not the standard one but that based on the
Fouks-Schubert model (Fouks \& Schubert \cite{Fouks95}) which is valid for
low contrasted ISOCAM LW observations (Coulais \& Abergel
\cite{Coulais00}). This was necessary since our LW observations were
preceded by short-wavelength (SW) measurements (which could not be used), 
which made more difficult to get the history of the LW detector.

\begin{table}
\caption{ISO journal of observations.\label{isoobs}}
\begin{tabular}{lcl}
\hline
Name & ISO\_id & Observing date\\
 & CAM/P2/C100\&C200 & \\
\hline
PKS 0048$-$097  & 38701401/02/03  & 1996 Dec 8\\
PKS 0118$-$272  & 74400104/05/06  & 1997 Nov 28\\
PKS 0138$-$097  & 76301007/08/09  & 1997 Dec 17\\
PKS 0426$-$380  & 69702410/11/12  & 1997 Oct 12\\
S5 0454+844     & 68700413/14/15  & 1997 Oct 2\\
PKS 0537$-$441  & 70201116/17/18  & 1997 Oct 18\\
PKS 0735+178    & 72301882/83/84  & 1997 Nov 8\\
PKS 1144$-$379  & 25400637/38/39  & 1996 Jul 27\\
OQ 530          & 40200746/47/48  & 1996 Dec 22\\
PKS 1519$-$273  & 43401152/53/54  & 1997 Jan 23\\
PKS 1749+701    & 39901458/59/60  & 1996 Dec 19\\
PKS 1749+096    & 82200287/88/89  & 1998 Feb 14\\
S5 1803$+$784   & 15100861/62/63  & 1996 Apr 16\\ 
4C 56.27        & 15100964/65/66  & 1996 Apr 16\\
PKS 2131$-$021  & 38601073/74/75  & 1996 Dec 6\\
PKS 2240$-$260  & 16402576/77/78  & 1996 Apr 29\\
PKS 2254+074    & 37900579/80/81  & 1996 Nov 29\\
\hline
\end{tabular}
\end{table}

The source flux was derived by performing aperture photometry with a radius
which contained $\sim 90\%$ of the flux for each filter, namely 2 pixels
for LW2, 3 pixels for LW7, and 4 pixels for LW3, correcting for the missing
flux using the CIA point spread function (PSF) library. The background was
derived in a four pixel wide annulus whose inner radius was two pixels
larger than the extraction radius for most sources (see below). The
aperture centre was determined in an iterative way by first placing it on
the brightest pixel and then by moving it in 1/4 pixel steps around that
position until the largest value was reached. The flux uncertainty was
estimated by adding in quadrature the root mean squared (rms) of the image
within the extraction radius and the $1\sigma$ uncertainty of the
background estimate. Systematic errors of ISOCAM photometry are $\sim 15\%$
for point sources (Blommaert et al. \cite{Blommaert03}), which are 
typically below our uncertainties. 

\begin{table}
\caption{ISOCAM fluxes$^a$ and slopes.\label{camflux}}
\begin{tabular}{llllr}
\hline
Name &LW2& \magg LW7&\magg LW3&$\alpha_{\nu}$~~~~ \\
     &[$6.7\mu$m] &\magg[$9.6\mu$m]&\magg[$14.3\mu$m]&  \\
\hline
PKS 0048$-$097&18.2$\pm$2.0 &\magg19.2$\pm$6.8 &\magg25.7$\pm$5.7 &0.4$\pm$0.4\\ 
PKS 0118$-$272&18.2$\pm$2.7 &\magg20.1$\pm$7.3 &\magg35.7$\pm$5.6 &0.9$\pm$0.3\\ 
PKS 0138$-$097&#9.1$\pm$2.2 &\magg11.9$\pm$5.5 &\magg18.9$\pm$5.2 &1.0$\pm$0.6\\ 
PKS 0426$-$380&#2.5$\pm$1.2 &$<$14.7           &\magg19.0$\pm$5.6 &2.7$\pm$0.7\\ 
S5 0454+844&#5.5$\pm$2.2 &$<$16.2           &\magg13.5$\pm$5.7 &1.2$\pm$0.9\\ 
PKS 0537$-$441&18.4$\pm$2.0 &\magg29.1$\pm$5.5 &\magg32.6$\pm$5.3 &0.8$\pm$0.2\\ 
PKS 0735+178&19.3$\pm$2.0 &\magg23.2$\pm$6.4 &\magg33.4$\pm$5.1 &0.7$\pm$0.3\\ 
PKS 1144$-$379&10.7$\pm$1.7 &\magg14.4$\pm$6.2 &\magg37.2$\pm$5.7 &1.6$\pm$0.3\\ 
OQ 530&21.0$\pm$2.6 &\magg32.8$\pm$5.3 &\magg44.1$\pm$5.2 &1.0$\pm$0.2\\ 
PKS 1519$-$273&#9.9$\pm$2.9 &\magg11.2$\pm$6.4 &$<$14.1           &0.4$\pm$1.6\\ 
PKS 1749+701&29.2$\pm$2.5 &\magg46.7$\pm$6.6 &\magg54.3$\pm$5.5 &0.8$\pm$0.2\\ 
PKS 1749+096&33.2$\pm$2.4 &\magg43.0$\pm$7.3 &\magg68.8$\pm$6.2 &1.0$\pm$0.2\\ 
S5 1803+784&29.3$\pm$2.0 &\magg33.0$\pm$5.7 &\magg66.8$\pm$5.1 &1.1$\pm$0.1\\ 
4C 56.27&#6.1$\pm$2.2 &\magg#8.8$\pm$3.7 &\magg#9.2$\pm$3.7 &0.5$\pm$0.6\\ 
PKS 2131$-$021 &12.1$\pm$2.3 &$<$17.6 &\magg10.7$\pm$4.3 &$-$0.2$\pm$0.5\\
PKS 2240$-$260 &#6.9$\pm$2.1 &\magg21.6$\pm$5.9 &\magg17.8$\pm$5.3&1.2$\pm$0.5\\ 
PKS 2254+074 &14.1$\pm$2.3 &\magg17.8$\pm$6.1 &\magg30.5$\pm$5.7 &1.0$\pm$0.4\\ 
\hline
\multicolumn{5}{l}{\footnotesize $^a$mJy} 
\end{tabular}
\end{table}

\subsubsection{Notes on individual sources} 

~~~~~{\bf PKS 0138--097}: faint source with a stronger source $\sim 6.5$
pixels away in the LW2 image; we then estimated the background in a two
(instead of four) pixel wide annulus.

{\bf PKS 0426--380}: faint source; to increase the signal to noise (S/N) we
used an extraction radius of 1 pixel for the LW2 filter, instead of our
default value of two. No significant detection in the LW7 filter.

{\bf S5 0454+844}: faint source with another source of comparable flux
$\sim 8$ pixels away in the LW2 image; we then estimated the background in
a three (instead of four) pixel wide annulus. No significant detection in
the LW7 filter. Faint source with another source of comparable flux $\sim
8$ pixels away in the LW3 image; we then estimated the background in a four
pixel wide annulus just outside the extraction circle.

{\bf PKS 0735+178}: another source of comparable flux $\sim 5.5$ pixels
away in the LW7 image; we then estimated the background in a four pixel
wide annulus whose inner radius was four (instead of two) pixels larger
than the extraction radius. Same for the LW3 filter (another source $\sim
6.5$ pixels away).

{\bf PKS 1519--273}: another source of comparable flux $\sim 7.5$ pixels
away in the LW7 image; we then estimated the background in a two (instead
of four) pixel wide annulus. No significant detection in the LW3 filter.

{\bf PKS 1749+701}: structure in the background of the LW3 image; we then
estimated the background in a region closer to the source (inner radius was
one pixel larger than the extraction radius and annulus was three pixels
wide).

{\bf 4C 56.27}: faint source; to increase the signal to noise (S/N) we used
an extraction radius of three pixels for the LW3 filter, instead of four.

{\bf PKS 2131--021}: faint source; no significant detection in the LW7
filter. To increase the signal to noise (S/N) we used an extraction radius
of three pixels for the LW3 filter, instead of four.

{\bf PKS 2240--260}: structure in the background of the LW2 image; we then
estimated the background in a region closer to the source (inner radius was
one pixel larger than the extraction radius).

\subsection{ISOPHOT}\label{isophot}

All sources were observed with ISOPHOT (Laureijs et al. \cite{Laureijs03})
in rectangular chopped mode, i.e., the line-of-sight was switched
periodically between the source and the background position, the latter
being placed at $180''$ from the source. For each source, data were taken
with the P2 detector at $25\mu$m, with the C100 detector at 65, 80 and
$100\mu$m, and with the C200 detector in the $120, 170$ and $180\mu$m
bands. Typical on-source exposure times were 128\,s (P2), 32-64\,s (C100),
and 32\,s (C200). A journal of the observations is presented in Tab.
\ref{isoobs}, which gives the name of the source, the corresponding
ISO\_id, and the observing date.

The data reduction was performed using the ISOPHOT Interactive Analysis
Software Package V10.0 (PIA, Gabriel et al. \cite{Gabriel98}). After
corrections for non-linearities of the integration ramps, an 8-point signal
pattern was created from each observation by overplotting and averaging the
basic blocks of the observation (the repeated background+source cycles; for
details see \'Abrah\'am et al. \cite{Abraham03}). The signals of the
patterns were transformed to a standard reset interval of 1/4 s, and an
orbital dependent dark current was subtracted. The signals were corrected
for non-linearities of the detectors by applying signal linearization
corrections. From the patterns a (source $-$ background) difference signal
was then extracted. In most cases the measured difference signal
underestimates the real signal due to short term detector transients.
Therefore, we applied the signal loss correction algorithm as implemented
in PIA. This correction is a function of the chopping frequency and the
measured difference signal. The flux calibration of the C100/C200
measurements was performed by comparison with the on-board fine calibration
source (FCS), which was also measured in chopped mode. At 25 $\mu$m the
detector's actual sensitivity could be reliably predicted from the orbital
position of the observation, and an orbital dependent default responsivity
was applied. Finally, the derived flux densities were corrected for the
finite size of the aperture by using the standard correction values.

In a chopped observation the slow detector baseline variations are expected
to be cancelled via the frequent alternation of the source and background
positions. In the case of very faint sources, however, the time per chopper
plateau was often rather long, and in many cases only two ON-OFF cycles
were performed. In these observations the baseline variations may not be
perfectly corrected via the chopping procedure, since the temporal
evolution of the baseline was only scarcely sampled. From a comprehensive
calibration study we found that in the case of P2 observations of faint
sources with only two chopper cycles the resulting fluxes were incorrect,
usually overestimating the true value. In order to cancel the signal
transient, we introduced a new correction at the beginning of the
processing scheme, by applying a transient correction developed
for staring (rather than chopped) P2 observations.

The staring transient correction normally works on signals at the Signal
per Ramp Data (SRD) level, but in the case of a chopped observation it has
to be applied before the pattern is created. Thus we modified the PIA
procedure "process\_erd\_to\_pattern.pro" and applied the transient
correction on each pair-wise signal before pattern creation. The agreement
with the fluxes predicted by extrapolating the ISOCAM fluxes was
significantly improved with the new correction.  We applied this algorithm
to all our P2 observations which had only two ON-OFF
cycles. We then processed further the measurements as normal chopped
observations. Notice that some residual problems with uncorrected P2 data
might remain (see \S~\ref{peak}).

The flux uncertainty at 25 $\mu$m was estimated to be of the order of
50\,mJy at faint flux levels. For the C100/C200 filters one important source
of systematic uncertainty may be related to the signal loss correction. In
addition, the actual responsivity of the detector had to be derived from
the accompanying FCS measurement, thereby introducing an additional
uncertainty factor. Finally, at long wavelengths ($\lambda \geq 100\mu$m) sky
confusion noise can be the dominant source of photometric error. 

Note that for the four sources for which ISOCAM detected a close-by,
relatively strong source, namely PKS 0138$-$097, S5 0454+844, PKS 0735+178,
and PKS 1519$-$273, some confusion is expected in the ISOPHOT band (see 
\S~\ref{peak}).

\begin{table*}
\caption{ISOPHOT fluxes$^a$. \label{photflux}}
\begin{tabular}{llllllll}
\hline
Name & $P2/25$ & $C100/C\_50$ & $C100/C\_70$ & $C100/C\_100$ &
     $C200/C\_120$ & $C200/C\_160$ & $C200/C\_180$ \\ 
     & [$25\mu$m] & [$65\mu$m] & $[80\mu$m] & [$100\mu$m] & [$120\mu$m]& 
     [$170\mu$m]& [$180\mu$m] \\
\hline
PKS 0048$-$097 &\magg 195$\pm$48  &\magg  218$\pm$115 &   $ <  $552 &	   $ <  $378 &	 $ <  $1032 &	 $ <  $1377 &   $ <  $#933\\ 
PKS 0118$-$272 &\magg #53$\pm$48  &\magg  196$\pm$69  &\magg   290$\pm$174  &\magg 150$\pm$132  &	 $ < # $948  &\magg   #331$\pm$249   &   $ <  $#822 \\ 
PKS 0138$-$097 &\magg 199$\pm$48  &\magg  317$\pm$76  &\magg   271$\pm$65   &   $ <  $549 & $ <  $1089 & $ <  $1524 &   $ <  $#981 \\ 
PKS 0426$-$380 &\magg #63$\pm$48  &\magg  126$\pm$60  &   $ <  $819 &  $ <  $180 &	 $ <  #$792  &\magg   #400$\pm$201   &   $ <  $#762 \\ 
S5 0454+844    &$ < $144&	  $ < $375&     $ <  $783 &$ <  $582 & $ <  $1746 &	 $ <  $3771 &   $ <  $6423 \\ 
PKS 0537$-$441 &\magg  #70$\pm$48 &\magg  249$\pm$60  &\magg   155$\pm$117  &$ <  $180 &\magg   #365$\pm$315   &\magg   #404$\pm$387   &   $ <  $#933 \\ 
PKS 0735+178   &\magg 225$\pm$48  &\magg  221$\pm$133 &      $ <  $366 &\magg 523$\pm$167  &$ <  $1197 &	 $ <  $#975  &  $ <  $1029 \\ 
PKS 1144$-$379 &\magg 115$\pm$48  &\magg  269$\pm$75  &\magg   259$\pm$61   &$ <  $309 &$ <  $1353 &	 $ <  $1566 &   $ <  $1857 \\ 
OQ 530         &\magg 139$\pm$48  &\magg  160$\pm$97  &\magg   183$\pm$60   &\magg 204$\pm$181  &$ < # $837  &\magg  #351$\pm$237   &\magg #337$\pm$251 \\ 
PKS 1519$-$273 &\magg 249$\pm$48  &\magg  312$\pm$77  &\magg   164$\pm$60   &\magg 137$\pm$115  &$ <  $1305 &	 $ <  $4215 &   $ <  $5025 \\ 
PKS 1749+701   &\magg 132$\pm$48  &$ <  $285&\magg   #76$\pm$60   &\magg 247$\pm$60   & $ <  #$846  &$ <  $2307 & $ <  $1725 \\ 
PKS 1749+096   &\magg 134$\pm$48  &\magg  503$\pm$75  &\magg   619$\pm$93   &\magg 346$\pm$102  &$ <  $1347 &	 $ <  $3495 &   $ <  $4146 \\ 
S5 1803+784    &\magg 162$\pm$48  &\magg  #75$\pm$60  &\magg   330$\pm$274  &\magg 312$\pm$60  &\magg  #410$\pm$305 &\magg #472$\pm$278  &\magg #507$\pm$300 \\
4C 56.27       &\magg  #63$\pm$48 &\magg    712$\pm$144 &      $ <  $429 &   $ <  $411 &$ <  $1182 &$ <  $1914 &\magg #459$\pm$416 \\ 
PKS 2131$-$021 &\magg 178$\pm$48  &$ <  $189&   $ <  $282 &\magg 170$\pm$60   & $ <  $1017 &	 $ <  $#927  &\magg #387$\pm$331 \\ 
PKS 2240$-$260 &\magg 139$\pm$48  &$ <  $195&   $ <  $213 &\magg 198$\pm$60   &	 $ <  #$909  &	 $ <  $1146 &   $ <  $1122 \\ 
PKS 2254+074   &\magg 190$\pm$48  &\magg  114$\pm$96  &\magg   379$\pm$115  &\magg 265$\pm$185  &$ <  $1017 &$ <  $1104 &$ <  $1134 \\ 

\hline
\multicolumn{8}{l}{\footnotesize $^a$mJy} 
\end{tabular}
\end{table*}

\subsection{Results} 

Our ISOCAM fluxes are reported in Tab. \ref{camflux}. When no significant
detection was obtained we give $3\sigma$ upper limits. Tab. \ref{camflux}
gives also the best fit spectral slopes (in frequency space) to the ISOCAM
data. These were derived by taking into account the errors on the fluxes
and by excluding non detections. The mean value is $\langle \alpha_{\nu}
\rangle = 0.95\pm0.15$ (median 1.0), while the weighted mean is $\langle
\alpha_{\nu} \rangle = 0.99\pm0.06$. This is consistent with expectations,
as a synchrotron peak frequency in the IR means $\nu f_{\nu} \approx {\rm
const}$ in the ISO bands, and therefore $f_{\nu} \propto \nu^{-1}$, as
indeed observed.

Our ISOPHOT fluxes are reported in Tab. \ref{photflux}. When no significant
detection was obtained, or the formal uncertainty was larger than the
measured flux density value, 3$\sigma$ upper limits are given.

\section{Discussion}\label{disc}

\subsection{Host galaxy contribution to the IR flux}

Before studying the spectral energy distributions (SEDs) of our sources it
is important to assess a possible contribution from the galaxy which hosts
the BL Lac. To this aim, we estimated the host galaxy flux in the IR band,
following the approach of Bertone et al.  (\cite{Bertone00}). A literature
search actually revealed that 13 out of our 17 BL Lacs are optically
unresolved, that is no host galaxy is detected (Pursimo et al.
\cite{Pursimo02}; Sbarufatti et al. \cite{Sbarufatti05}). For the remaining
four sources, we found host galaxy $R$ magnitudes in Pursimo et al.
(\cite{Pursimo02}) and converted them to $B$ magnitudes using the
evolutionary synthesis model of Dunlop et al. (\cite{Dunlop89}). We then
used the results of Mazzei \& DeZotti (\cite{Mazzei94}), who calculated the
flux ratio between the IRAS and the $B$ bands for a sample of 47
ellipticals, to estimate the galaxy contributions at 12, 25, 60, and 100
$\mu$m. These turned out to be {\it extremely} low. The maximum expected
contributions in fact, where around $1\%$, $0.01\%$, $0.1\%$, and $0.2\%$
of the observed fluxes at 12, 25, 60, and 100 $\mu$m respectively. We thus
conclude that for the BL Lacs in our sample the contribution of the host
galaxy to their ISO flux is completely negligible.

\subsection{Spectral energy distributions}

To constrain the synchrotron power peak and to address the relevance of our
ISO data in terms of emission processes in BL Lacs, we have assembled
multifrequency data for all our sources. We have also included {S5
2007+777}, a 1 Jy BL Lac at $z = 0.342$ with previously published ISOPHOT
data (Peng et al.  \cite{Peng00}). Since we are not interested in
characterizing the inverse Compton emission, we limited ourselves to
frequencies $\le 10^{16}$ Hz. The main source of information was NED.
Additional data were taken from the WMAP catalogue (Bennett et al.
\cite{Bennett03}), the Two Micron All Sky Survey (2MASS) (Cutri et
al. \cite{Cutri00}), the Guide Star Catalogue II\footnote{\tt
htpp://archive.stsci.edu/gsc/}, and the Sloan Digital Sky Survey
(SDSS). Therefore, most data are not simultaneous with our
observations. For more than half of our sources, however, we were able to
find nearly-simultaneous (typically within a month) radio observations in
the University of Michigan Radio Astronomy Observatory (UMRAO)
database\footnote{The Bologna group has also performed a radio monitoring
of some of our sources (Venturi et al.  \cite{Venturi01}). The UMRAO
observations, however, were always closer in time to our observing
dates.}. These are reported in Table \ref{radioobs}. Furthermore,
nearly-simultaneous optical data were available for PKS 0735+178 (Bai et
al. \cite{Bai99}), OQ 530 (Massaro et al. \cite{Massaro04}), and S5
1803+784 (Nesci et al. \cite{Nesci02}). As regards S5 2007+777, Peng et
al. (\cite{Peng00}) published simultaneous radio (seven frequencies),
ISOPHOT (60 and 100 $\mu$m), and optical ($R$ band) observations of this
object.

\begin{table*}
\caption{Nearly-simultaneous radio observations.\label{radioobs}}
\begin{tabular}{lclclcl}
\hline
Name &F$_{\rm 4.8GHz}$&Observing date&F$_{\rm 8.0GHz}$&Observing date&F$_{\rm 14
.5GHz}$&Observing date\\
     &              (Jy)&              &              (Jy)&              &
 (Jy)& \\
\hline
PKS 0048$-$097&$1.58\pm0.06$&1996 Dec 13 &$1.96\pm0.03$&1996 Dec 9 &$2.12\pm0.05$&1996 Dec 11\\
PKS 0735+178  &$1.06\pm0.04$&1997 Dec 2  &$0.69\pm0.16$&1997 Nov 24&$0.94\pm0.11$&1997 Dec 10\\
OQ 530           & ... & ...&$0.56\pm0.15$&1997 Jan 6 &$0.64\pm0.03$&1996 Dec 30     \\
PKS 1749+701  &$0.63\pm0.04$&1996 Dec 13& ... & ...&$0.86\pm0.03$&1996 Dec 19  \\
PKS 1749+096  &$2.26\pm0.03$&1998 Feb 18&$3.40\pm0.08$&1998 Feb 17&$4.02\pm0.09$&1998 Feb 4 \\
S5 1803+784&$2.91\pm0.04$&1996 Apr 3&$2.90\pm0.19$&1996 Apr 23&$3.18\pm0.06$
&1996 Apr 11\\
4C 56.27      &$1.79\pm0.04$&1996 Apr 16&$2.34\pm0.06$&1996 Apr 24 & ... & ...\\
PKS 2131$-$021& ... & ...&$1.67\pm0.02$&1997 Jan 6&$1.56\pm0.04$&1996 Dec 19 \\
PKS 2254+074  & ... & ...& ... & ...&$0.48\pm0.02$&1996 Dec 19 \\
\hline
\end{tabular}
\end{table*}

The SEDs for our sources are shown in Fig. \ref{seds}, where bigger, darker
symbols and upper limits indicate ISO data and the nearly-simultaneous
radio and optical data, and smaller, lighter symbols represent
non-simultaneous data collected from the literature.

\subsubsection{Synchrotron peak frequencies}\label{peak}

We determined $\nu_{\rm peak}$ for our sources by applying an homogeneous
synchrotron -- inverse self-Compton (SSC) model in the log $\nu$ -- log
$\nu f_{\nu}$ plane to the SEDs. The SSC model was adapted from Tavecchio
et al. (\cite{Tav98}) and assumes that radiation is produced by a
population of relativistic electrons emitting synchrotron radiation in a
single zone of a jet that is moving at relativistic speed and at a small
angle to the line of sight (see Giommi et al. \cite{Giommi02} and Padovani
et al. \cite{Padovani03} for previous applications of this model and more
details). Padovani et al. (\cite{Padovani03}) make the point that, despite
the fact that we cannot fully constrain the model parameters, $\nu_{\rm
peak}$, a combination of magnetic field strength, Doppler factor, and
electron break energy, is relatively well constrained even for relatively
large variations of these parameters, which affect more strongly the
high-energy/inverse Compton part of the SED.

Some previous papers have derived $\nu_{\rm peak}$ by fitting analytical
functions, such as a parabola or a third-degree polynomial, to the SED of
blazars (e.g., Sambruna et al. \cite{Sambruna96}; Fossati et al.
\cite{Fossati98}). We believe that our approach, although more complex and
time consuming, is more robust especially when dealing with sparsely
sampled SEDs, as we are guided by physics rather than just analytical
fitting.

During our fitting procedure we adhered to the following guidelines: 1.
more weight was given to our nearly-simultaneous data; 2. more weight was
given to ISOCAM data, which have smaller error bars; 3. clear outliers were
not taken into account; in case of ISOPHOT data, these indicate confusion
(see \S~\ref{isophot}) and/or some residual calibration issues; 4. only
data points with $\nu \ga 3 \times 10^9$ Hz were included. Our derived
rest-frame $\nu_{\rm peak}$ values are given in Tab. \ref{peaks}. Based on
our experience, we believe that most of our values are determined to better
than 0.5 dex. Exceptions to this rule include PKS 1519$-$273, PKS
2131$-$021, and PKS 2254$+$074, whose SEDs are sparse and/or whose ISO
continuum shape is somewhat irregular (see Fig. \ref{seds}).

The $\nu_{\rm peak}$ distribution is very narrow, with 10/18 sources having
values in the range $1 - 3 \times 10^{13}$ Hz. The mean value is $\langle
\log \nu_{\rm peak} \rangle = 13.0\pm0.1$ (median 13.0), that is $\sim
30\mu$m. Given our set of simultaneous IR data, these represent the
best determinations available of the synchrotron peak frequencies for the 1
Jy BL Lac sample.

\begin{table}
\caption{Rest-frame synchrotron peak frequencies and wavelenghts.\label{peaks}}
\begin{tabular}{llr}
\hline
Name & $\nu_{\rm peak}$ & $\lambda_{\rm peak}$\\
 & $10^{13}$ Hz & $\mu$m\\
\hline
PKS 0048$-$097  &#0.3& 100 \\
PKS 0118$-$272  &10&     3\\
PKS 0138$-$097  &#0.3& 100 \\
PKS 0426$-$380  &#0.5& 60\\ 
S5 0454+844     &#0.5& 60\\
PKS 0537$-$441  &#0.1&  300\\
PKS 0735+178    &#1&  30\\
PKS 1144$-$379  &#1&  30\\
OQ 530          &#3&  10\\
PKS 1519$-$273  &#1&  30\\
PKS 1749+701    &#3&  10\\
PKS 1749+096    &#0.3& 100 \\
S5 1803$+$784   &#1&   30\\
4C 56.27        &#0.5& 60\\
S5 2007+777$^a$ &#1&  30\\
PKS 2131$-$021  &#2&  15\\
PKS 2240$-$260  &#2&  15\\
PKS 2254+074    &#2&  15\\
\hline
\multicolumn{3}{l}{\footnotesize $^a$ISOPHOT data published by Peng et
al. (\cite{Peng00}).}
\end{tabular}
\end{table}

\subsubsection{Synchrotron peak variability in IR band} 

Inspection of Fig. \ref{seds} shows that quite a few of our sources have a
nearly-simultaneous SED around the peak which is quite different from the
one previously available using non-simultaneous data from the literature.
The most obvious cases include PKS 0048$-$097, PKS 0537$-$441, PKS
0735$+$178, OQ 530, 4C 56.27, where $\nu_{\rm peak}$ could have changed by
over one order of magnitude. This impression is confirmed by a comparison
with the $\nu_{\rm peak}$ values derived by Sambruna et al.
(\cite{Sambruna96}) for these objects, which are larger than ours by
factors $\sim 20$, 120, 10, 2, and 8 respectively. This may be indicative
of large changes in the location of the synchrotron peak, as observed in
the case of the high-energy peaked BL Lac MKN 501 (e.g., Pian et
al. \cite{Pian98}; Massaro et al. \cite{Massaro06}). Alternatively,
$\nu_{\rm peak}$ could have stayed constant, and the large optical
variability would be due to a shift in normalization of the whole SED,
implying a correspondingly high change in the IR flux. Simultaneous 
optical-IR monitoring is required to be able to distinguish
between the two possibilities.

\subsection{Astrophysical relevance}

A precise measurement of $\nu_{\rm peak}$ can obviously characterize
synchrotron emission from blazars. Furthermore, its value and that of the
spectral curvature around the maximum emission can constrain particle
acceleration mechanisms in these sources (e.g., Massaro et al.
\cite{Massaro06}). This is also extremely relevant for two up-coming
$\gamma$-ray missions, that is AGILE and GLAST. These observatories, in
fact, will sample inverse Compton emission above the peak for LBL and
flat-spectrum radio quasars. Moreover, the large synchrotron peak
variability possibly present in a few of our sources will also be important
to constrain the link between the two main emission processes in blazars
and the source of the inverse Compton seed photons (e.g., Sokolov et
al. \cite{Sokolov04}).

\subsection{Previous ISO observations of BL Lacs}

To the best of our knowledge, only three ISO observations of BL Lacs have
been published. Beside the already mentioned S5 2007+777, these include 
PKS 2155$-$304 (ISOCAM and ISOPHOT; Bertone et al. \cite{Bertone00}) 
and Mrk 180 (ISOPHOT, one detection at $90\mu$m; Anton et al. 
\cite{Anton04}), both of the HBL type. 

\section{Conclusions}

We have presented new, simultaneous ISOCAM and ISOPHOT observations
covering the $7 - 200\mu$m range for seventeen (out of 34) BL Lacertae
objects belonging to the 1 Jy sample, the prototype radio-selected
(low-energy peaked) sample. These data open up a new window on the spectral
energy distributions of low-energy peaked BL Lacs and fill a glaring gap in
their far-infrared spectra. They also represent a sevenfold (eighteenfold)
increase in the number of published ISO results for BL Lacs (LBL).

Our main results can be summarized as follows:

\begin{enumerate}

\item Our detection rate is good: 100\%, 82\%, and 94\% at 6.7 (LW2), 9.6
(LW7), and 14.3 (LW3) $\mu$m (ISOCAM), and 94\%, $\sim 65\%$, and $\sim
20\%$ at 25 (P2), $65 - 100$ (C100), and $120 - 180$ (C200) $\mu$m
(ISOPHOT) respectively. To deal with the effect of chopping on our
relatively faint sources, which would lead to a flux overestimate at 25
$\mu$m, we have also employed a novel correction in the ISOPHOT data
reduction;

\item We have built the spectral energy distributions of our sources, with
the addition of a previously published one, by using our new ISO data,
complemented by nearly-simultaneous radio data for about half the sample
(and optical data for four sources), and other non-simultaneous
multi-frequency data. We have then derived the synchrotron peak frequency,
$\nu_{\rm peak}$. This turned out to have a very narrow distribution, with
$\sim 60\%$ of the sources having values in the range $1 - 3 \times
10^{13}$ Hz, centered at $\sim 10^{13}$ Hz ($\sim 30\mu$m). These represent
the best determinations available of the synchrotron peak frequencies for
low-energy peaked BL Lacs;

\item A comparison with previous such estimates, based on non-simultaneous
optical and near infrared data, may indicate strong $\nu_{\rm peak}$
variations in a number of sources, possibly associated with large flares as
observed in the high-energy peaked BL Lac MKN 501.

\end{enumerate}

These results, apart from characterizing synchrotron emission in blazars,
can constrain particle acceleration mechanisms and are also important for
the planning and data interpretation of up-coming new $\gamma$-ray
missions.

\begin{acknowledgements}

PP thanks the European Space Astronomy Centre (ESAC; formerly known as
VILSPA) for its hospitality at the start of the data reduction process,
Rosario Lorente for her expert ISOCAM assistance, and Piero Rosati for IDL
help. We acknowledge the support of Paul Barr, Tommaso Maccacaro, Gianpiero
Tagliaferri, and Anna Wolter at an early stage of this project. The ISOCAM
data presented in this paper were analysed using CIA, a joint development
by the ESA Astrophysics Division and the ISOCAM Consortium. The ISOCAM
Consortium is led by the ISOCAM PI, C. Cesarsky. The ISOPHOT data presented
in this paper were reduced using PIA, which is a joint development by the
ESA Astrophysics Division and the ISOPHOT consortium, with the
collaboration of the Infrared Analysis and Processing Center (IPAC) and the
Instituto de Astrof\'isica de Canarias (IAC). This research has made use of
data from the University of Michigan Radio Astronomy Observatory which is
supported by funds from the University of Michigan, of the NASA/IPAC
Extragalactic Database (NED), which is operated by the Jet Propulsion
Laboratory, California Institute of Technology, under contract with the
National Aeronautics and Space Administration, of data products from the
Two Micron All Sky Survey, which is a joint project of the University of
Massachusetts and the Infrared Processing and Analysis Center/California
Institute of Technology, funded by the National Aeronautics and Space
Administration and the National Science Foundation, of Guide Star Catalogue
II data, and of Sloan Digital Sky Survey (SDSS) data.  Funding for the SDSS
and SDSS-II has been provided by the Alfred P. Sloan Foundation, the
Participating Institutions, the National Science Foundation, the
U.S. Department of Energy, the National Aeronautics and Space
Administration, the Japanese Monbukagakusho, the Max Planck Society, and
the Higher Education Funding Council for England. The SDSS Web Site is {\tt
http://www.sdss.org/}.

\end{acknowledgements}

\begin{figure*}[t]
\centering
\includegraphics{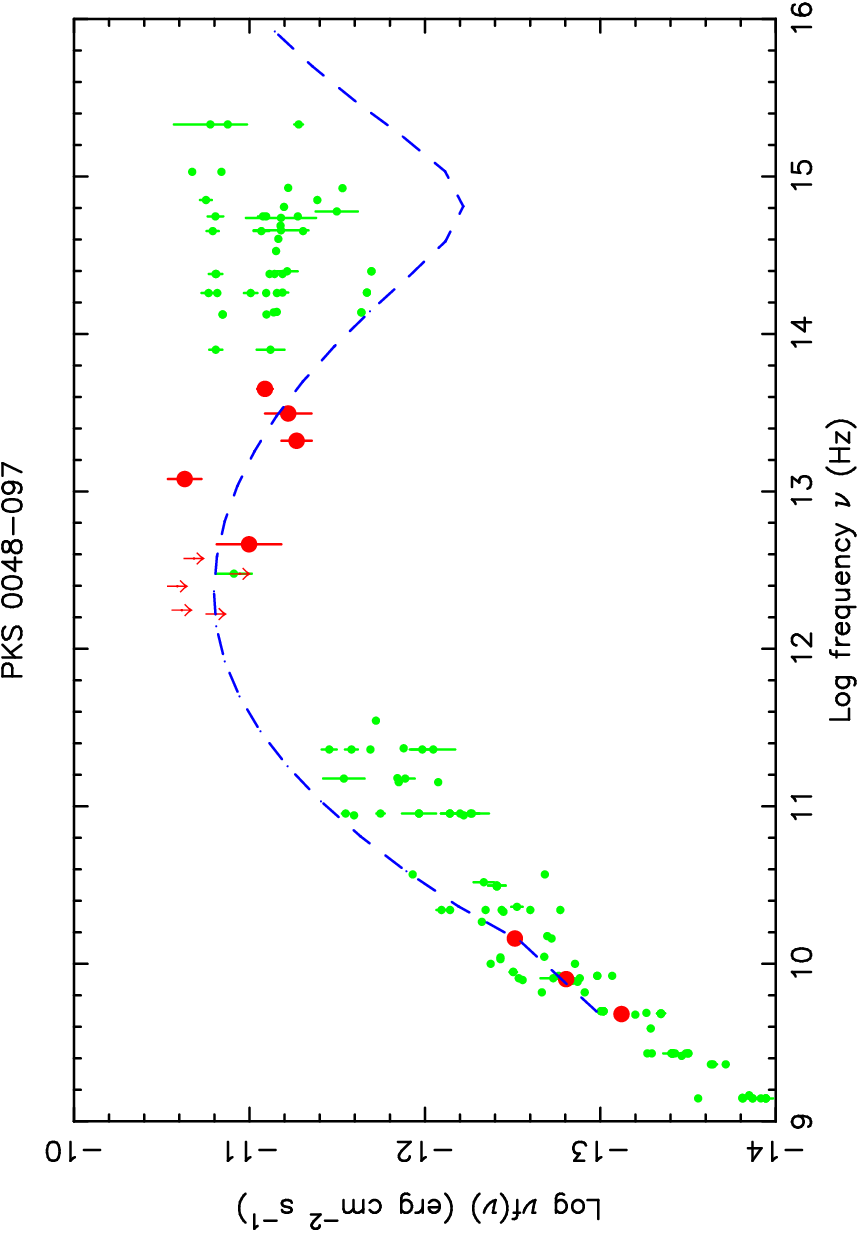} 
\includegraphics{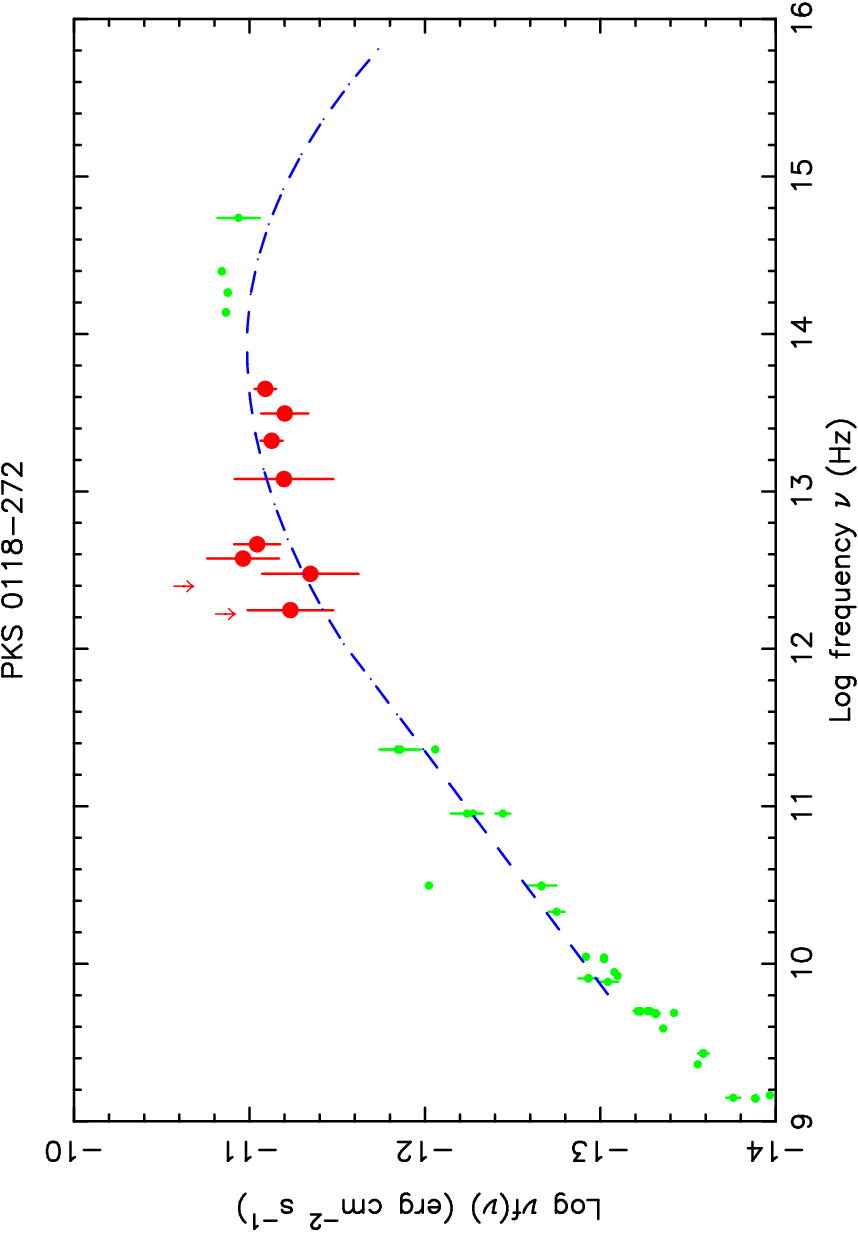} 
\includegraphics{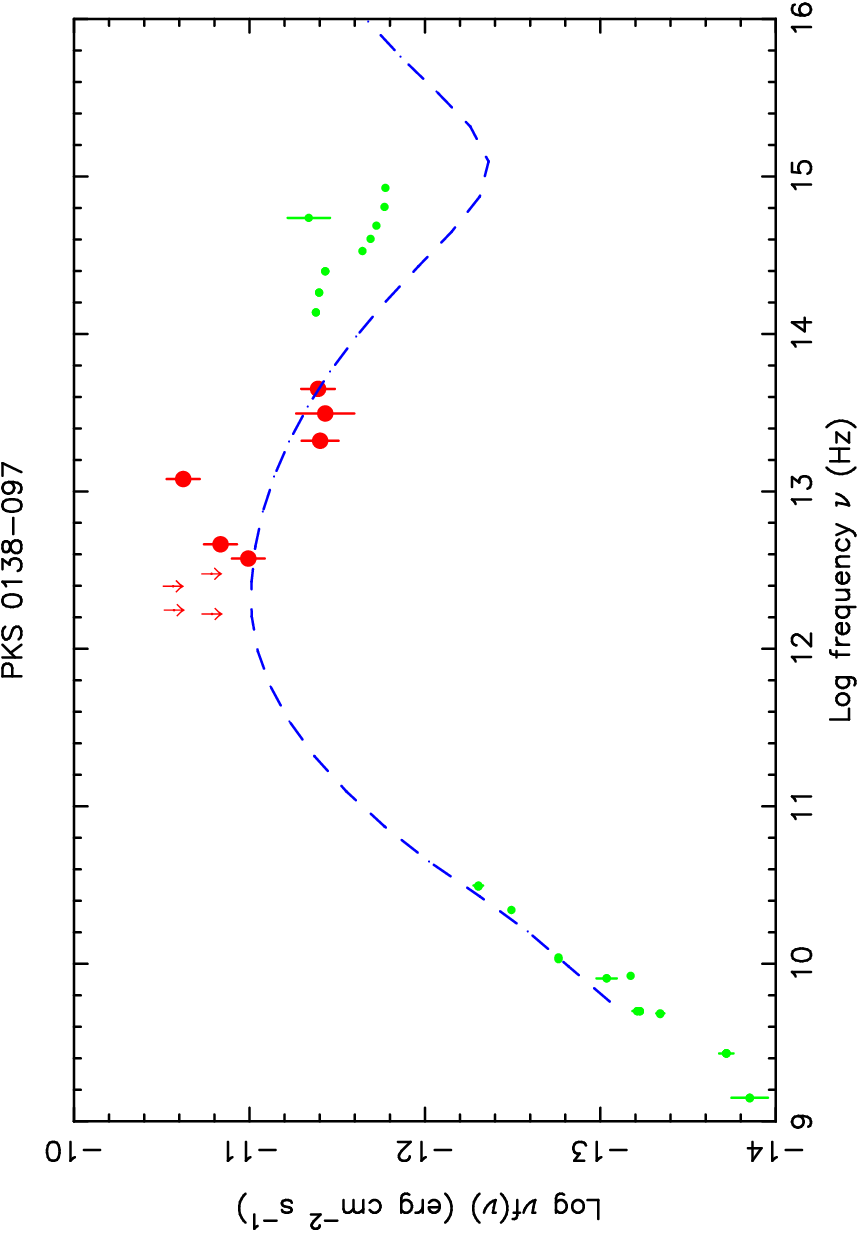} 
\includegraphics{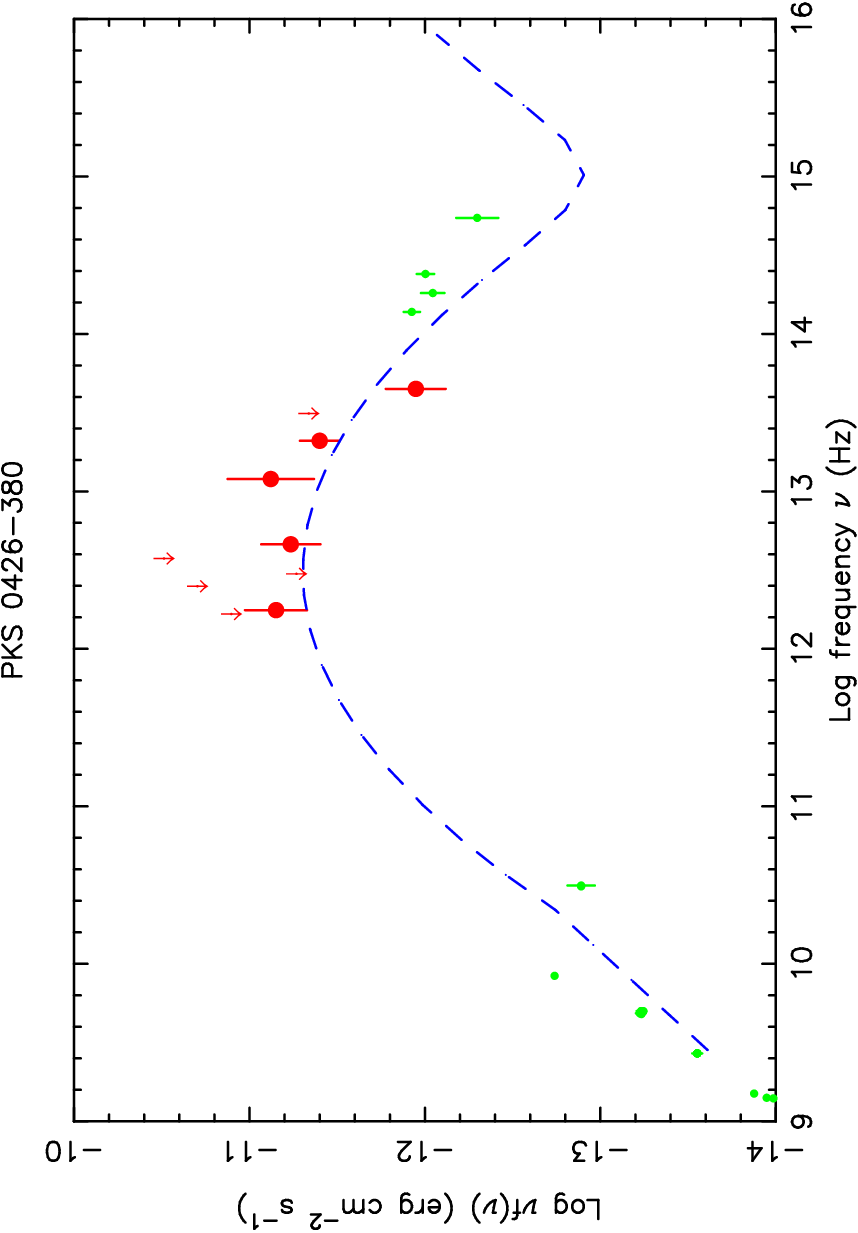} 
\vspace{22.5cm} 
\caption[t]{The spectral energy distributions for our sources. Bigger,
darker symbols indicate ISO data and the nearly-simultaneous radio (and
optical in the case of PKS 0735+178, OQ 530, S5 1803+784, and S5 2007+777)
data (arrows denote ISO upper limits), while smaller, lighter symbols
represent non-simultaneous data collected from the literature. The dashed
line is our synchrotron model.}\label{seds}
\end{figure*} 

\setcounter{figure}{0}
\begin{figure*}[t]
\centering
\includegraphics{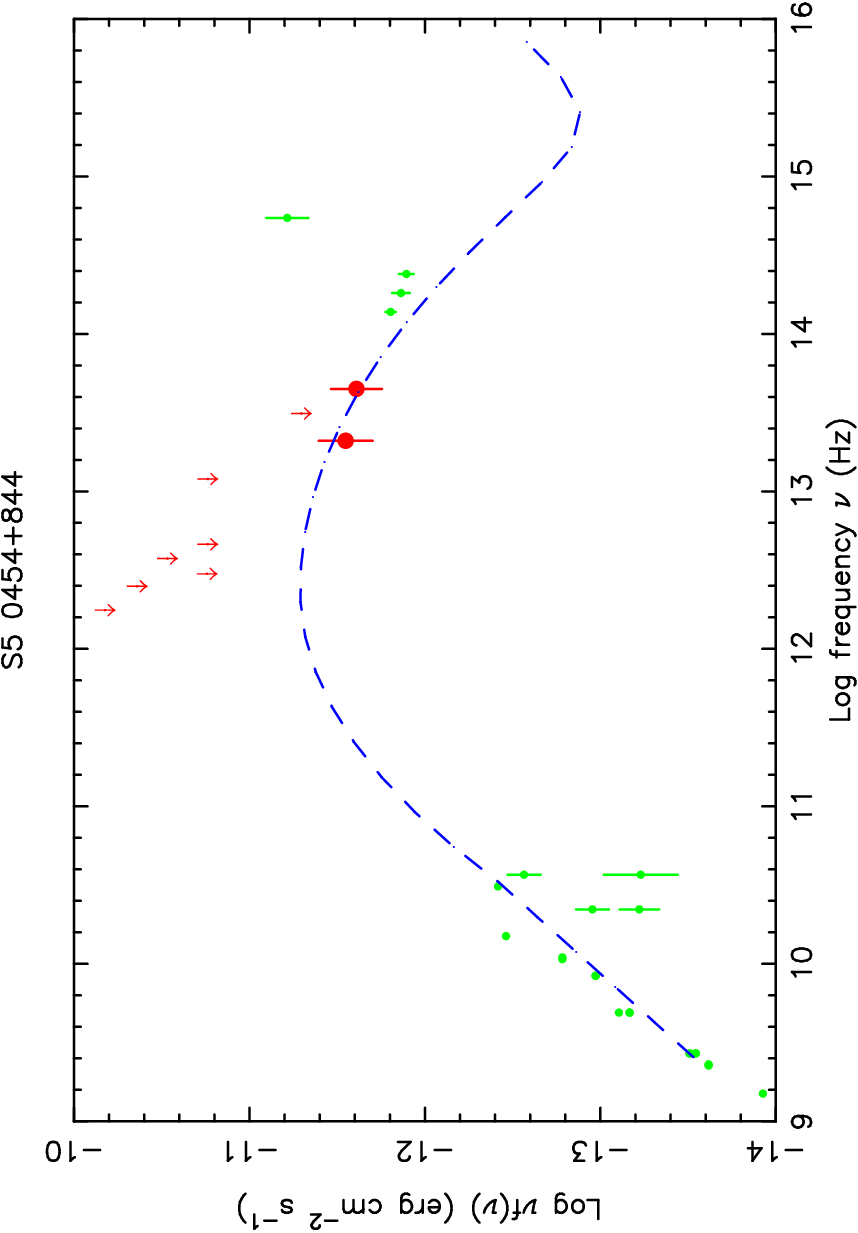} 
\includegraphics{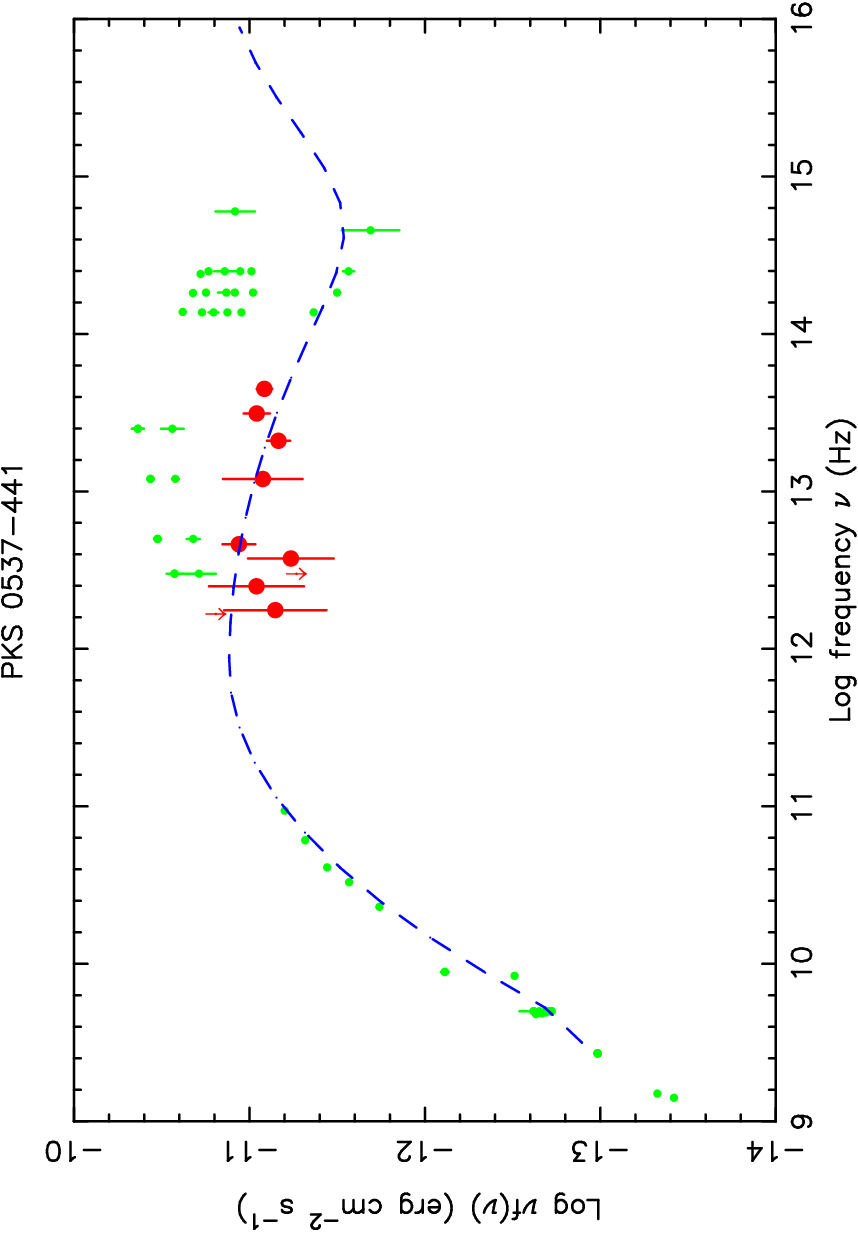} 
\includegraphics{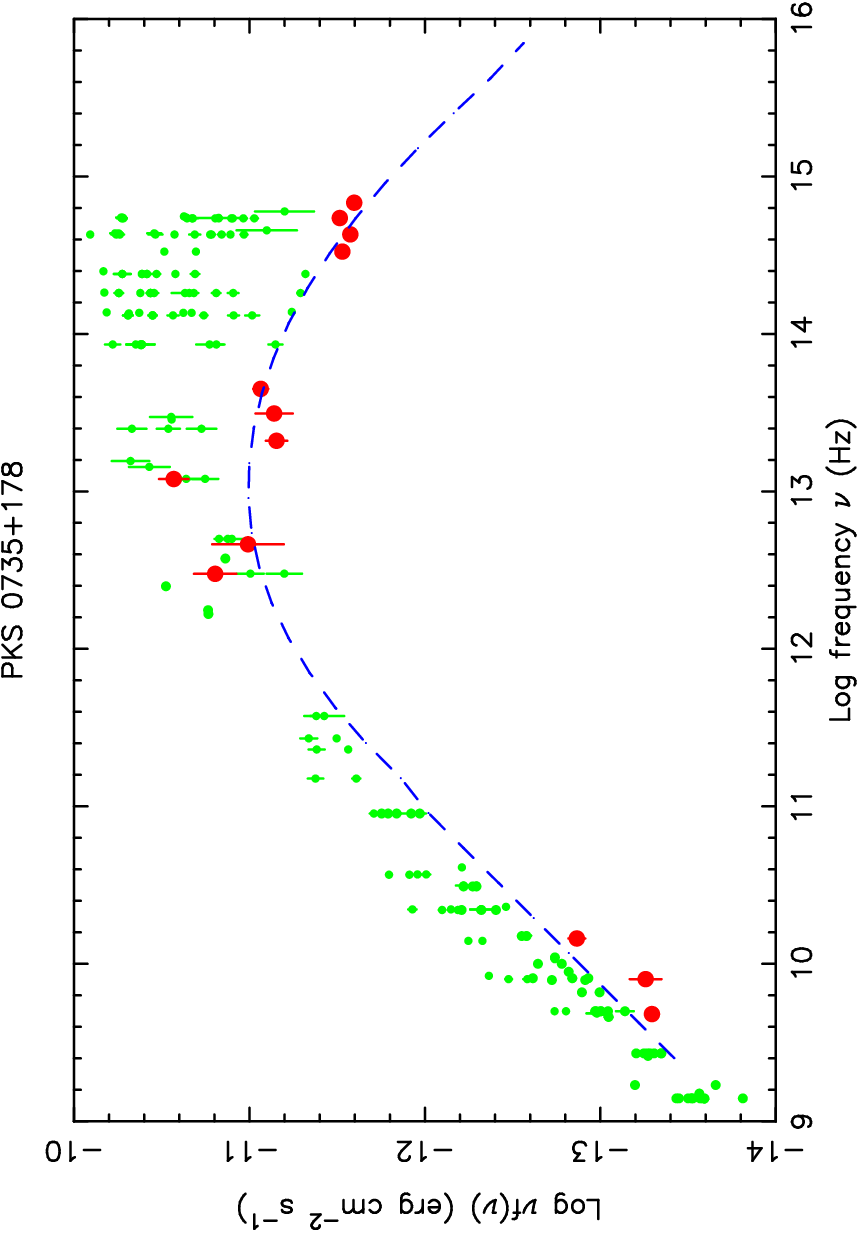} 
\includegraphics{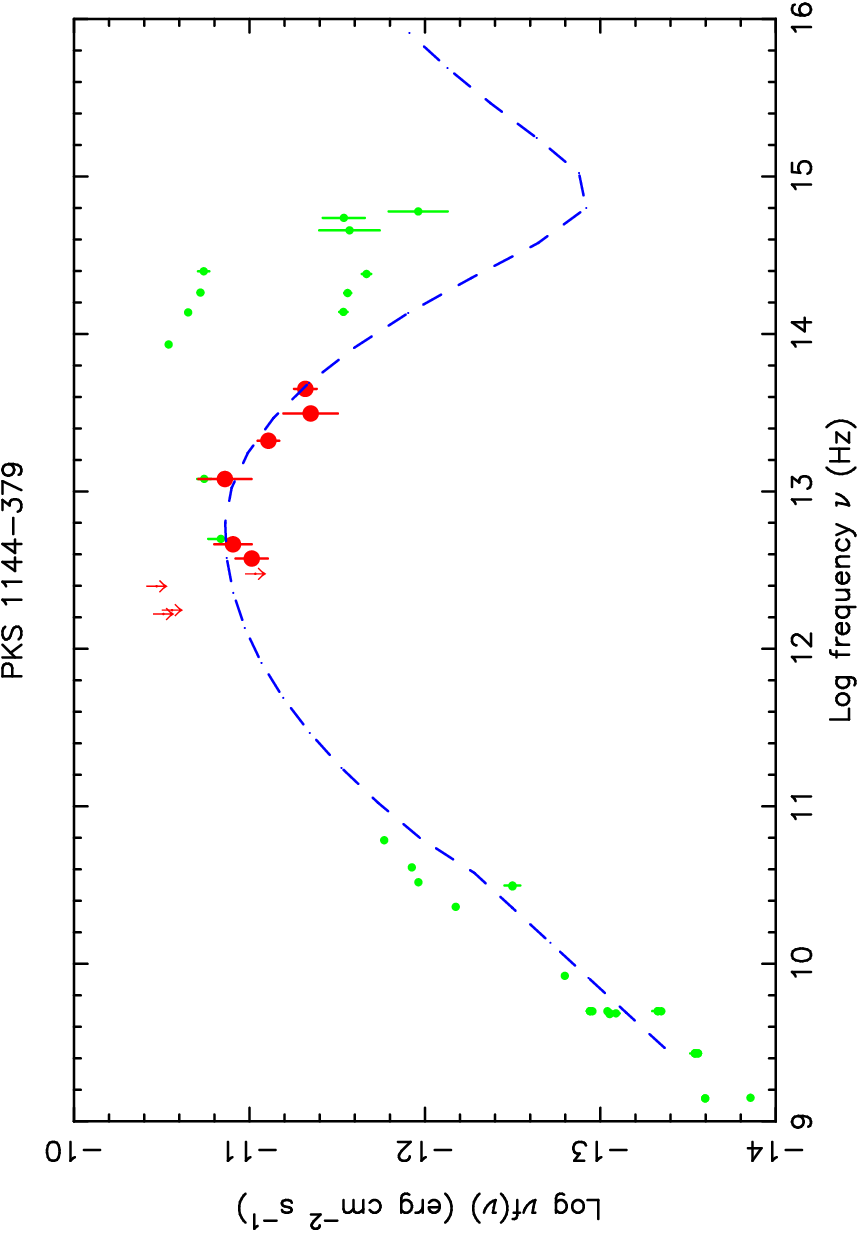} 
\vspace{22.5cm} 
\caption[t]{(continued) The spectral energy distributions for our
sources. Bigger, darker symbols indicate ISO data and the
nearly-simultaneous radio (and optical in the case of PKS 0735+178, OQ 530,
S5 1803+784, and S5 2007+777) data (arrows denote ISO upper limits), while
smaller, lighter symbols represent non-simultaneous data collected from the
literature. The dashed line is our synchrotron model.}\label{seds}
\end{figure*} 

\setcounter{figure}{0}
\begin{figure*}[t]
\centering
\includegraphics{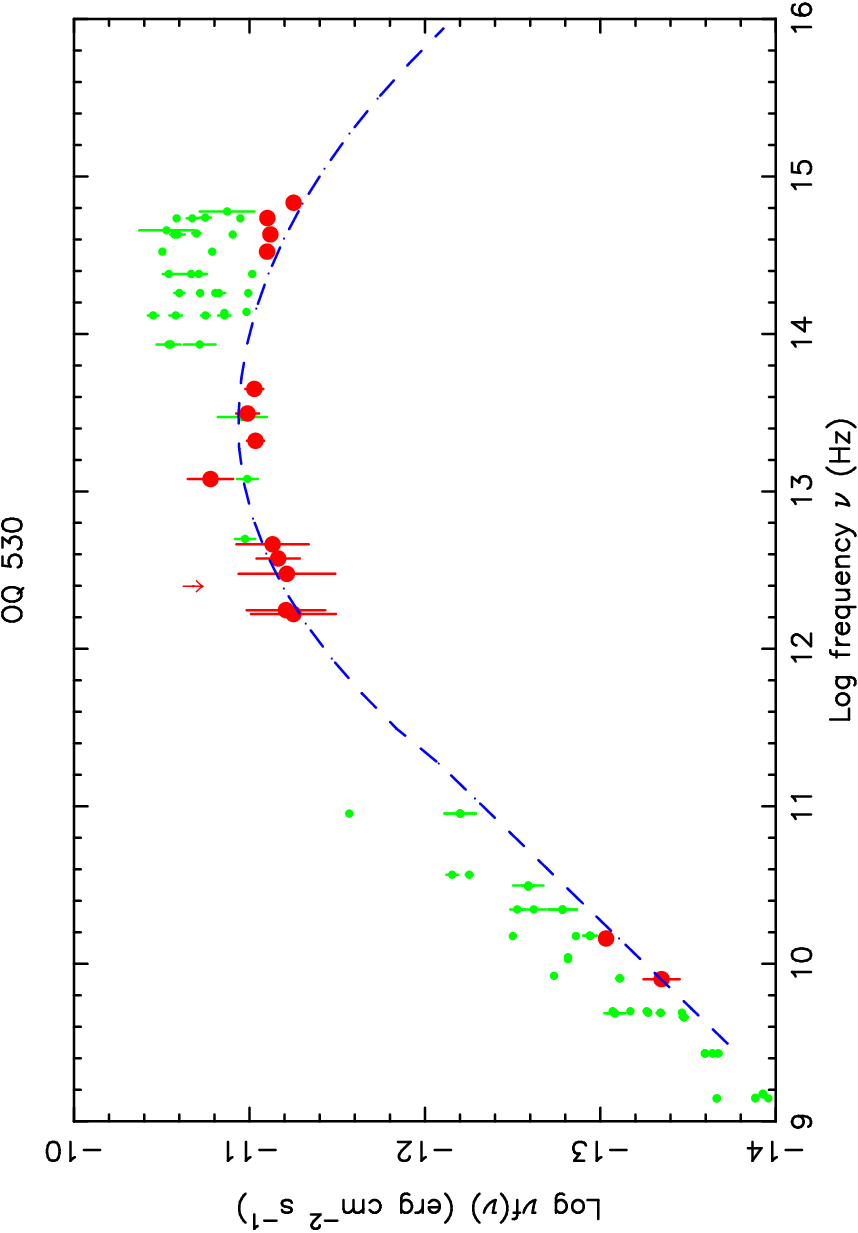} 
\includegraphics{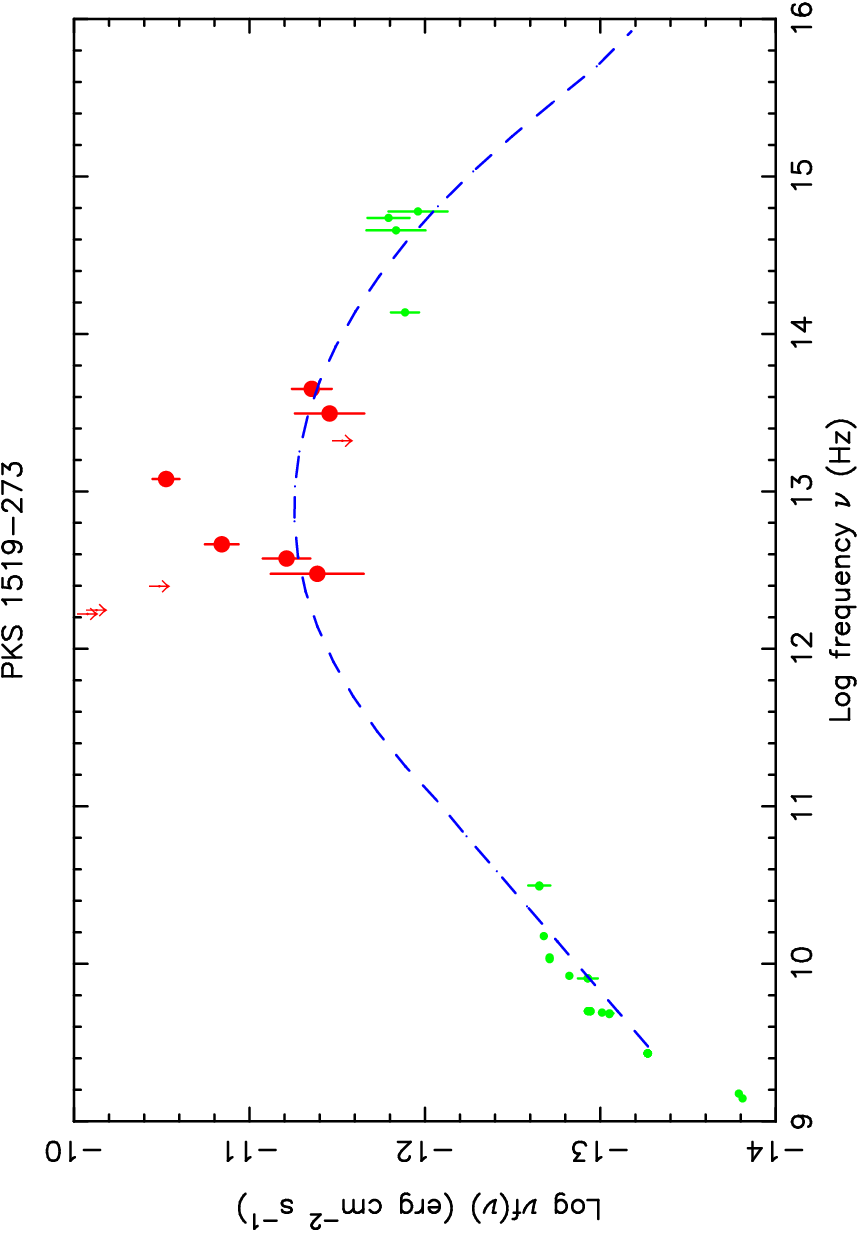} 
\includegraphics{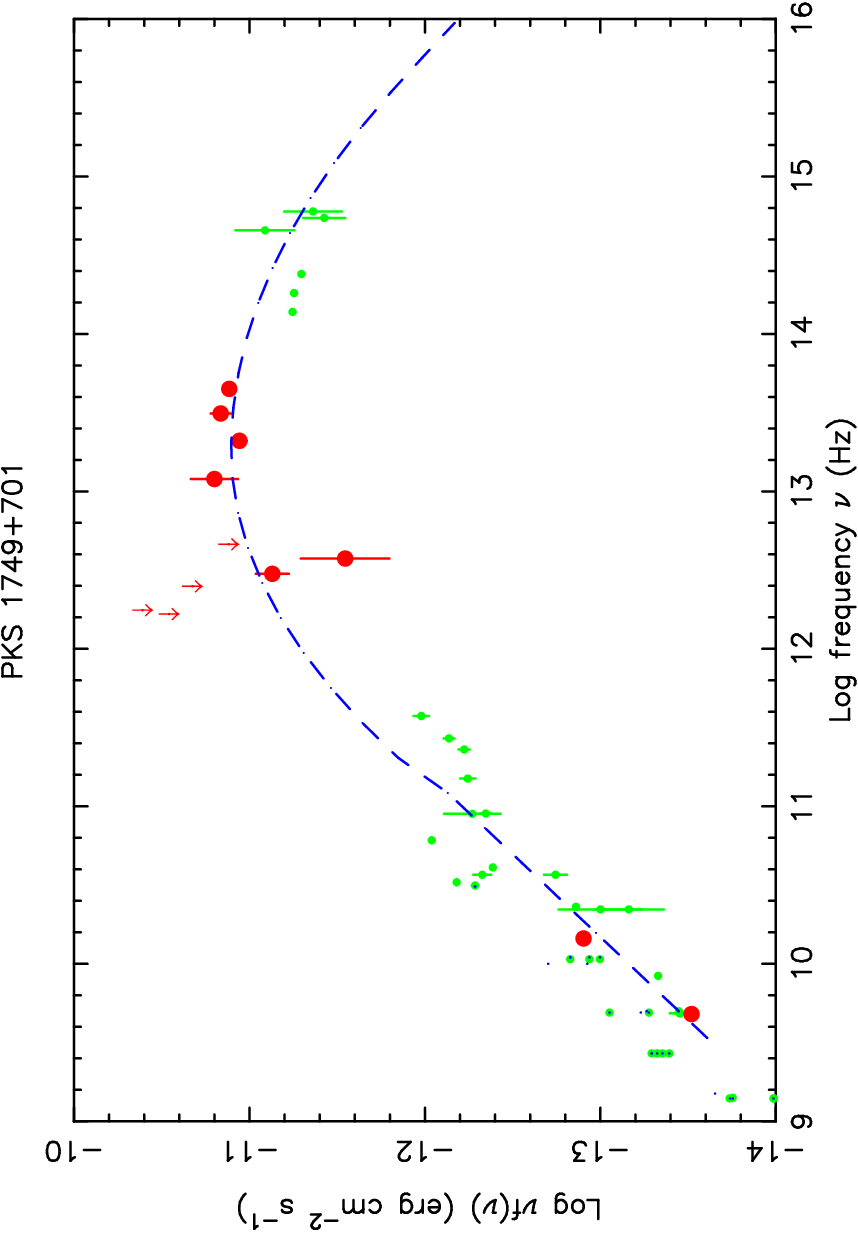} 
\includegraphics{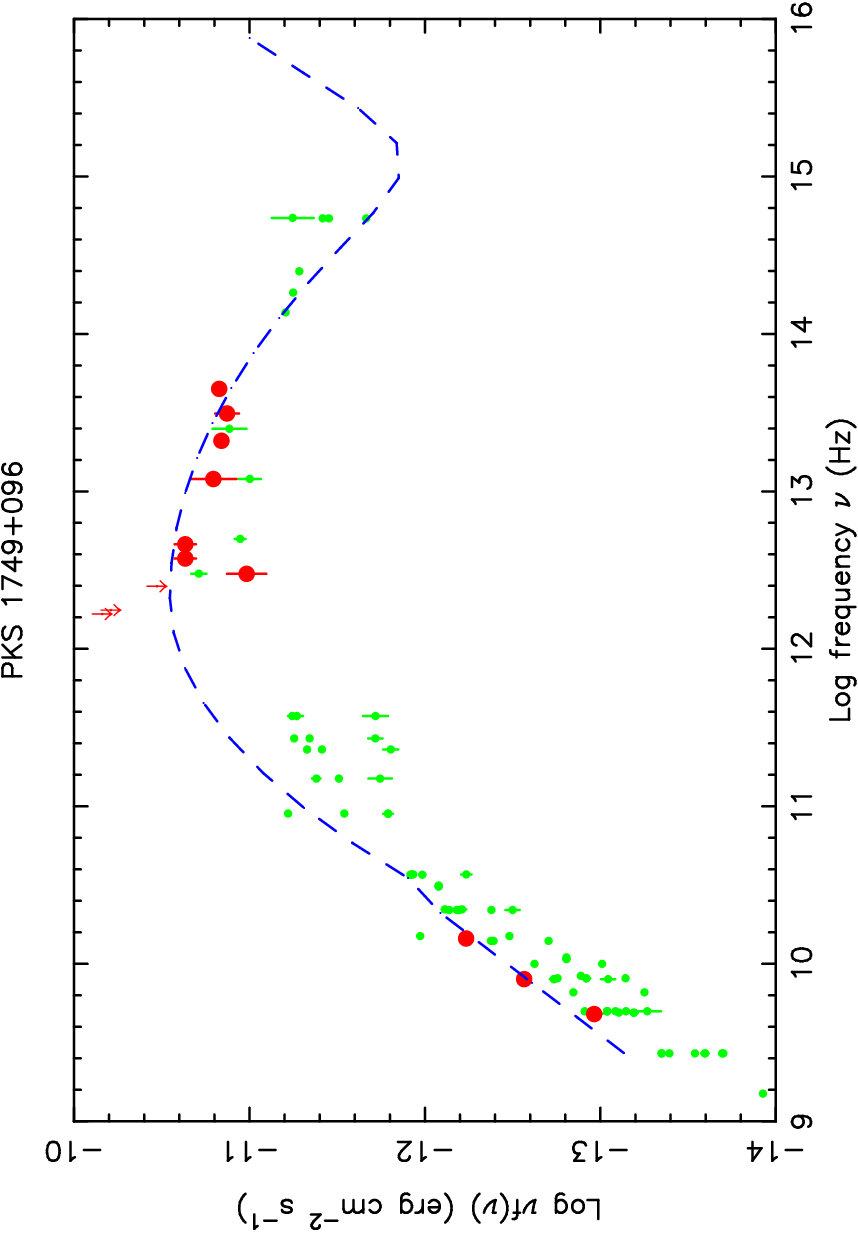} 
\vspace{22.5cm} 
\caption[t]{(continued) The spectral energy distributions for our
sources. Bigger, darker symbols indicate ISO data and the
nearly-simultaneous radio (and optical in the case of PKS 0735+178, OQ 530,
S5 1803+784, and S5 2007+777) data (arrows denote ISO upper limits), while
smaller, lighter symbols represent non-simultaneous data collected from the
literature. The dashed line is our synchrotron model.}\label{seds}
\end{figure*} 

\setcounter{figure}{0}
\begin{figure*}[t]
\centering
\includegraphics{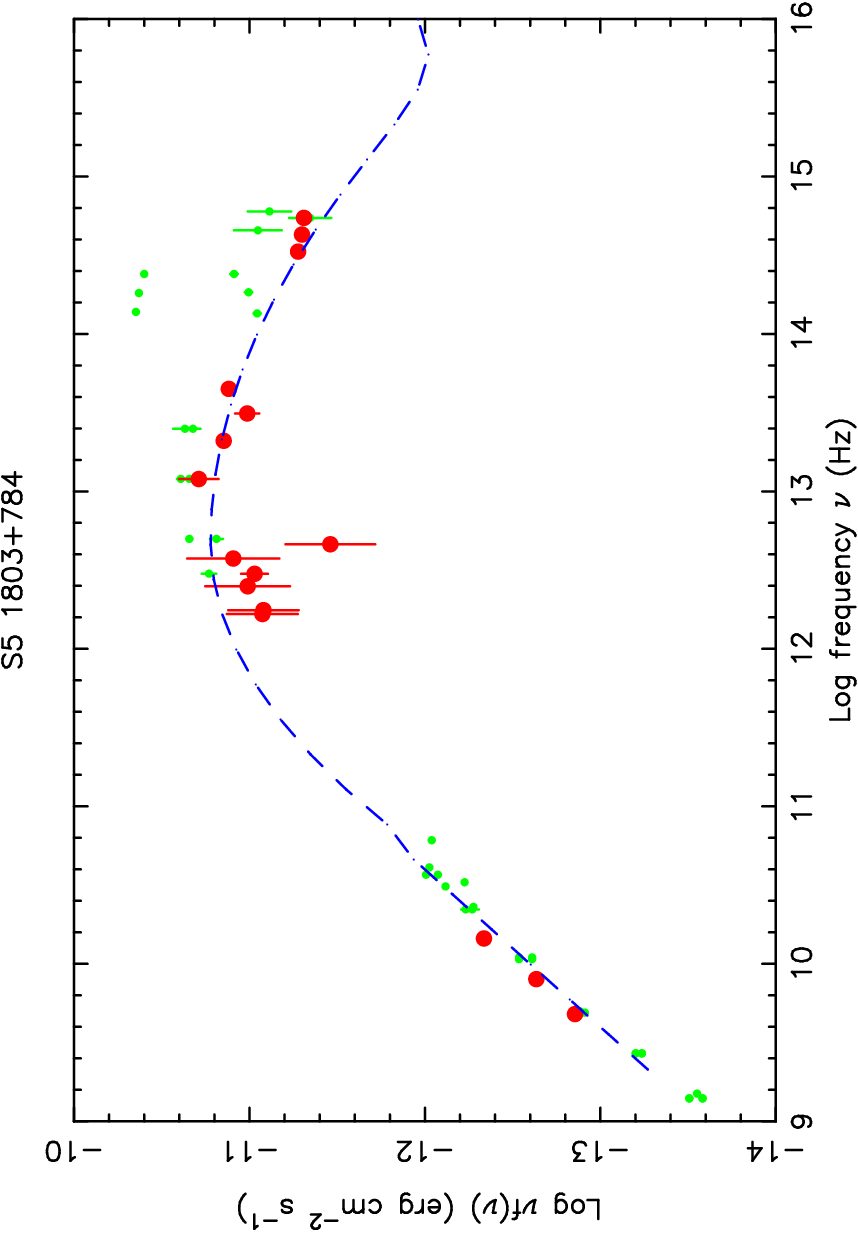} 
\includegraphics{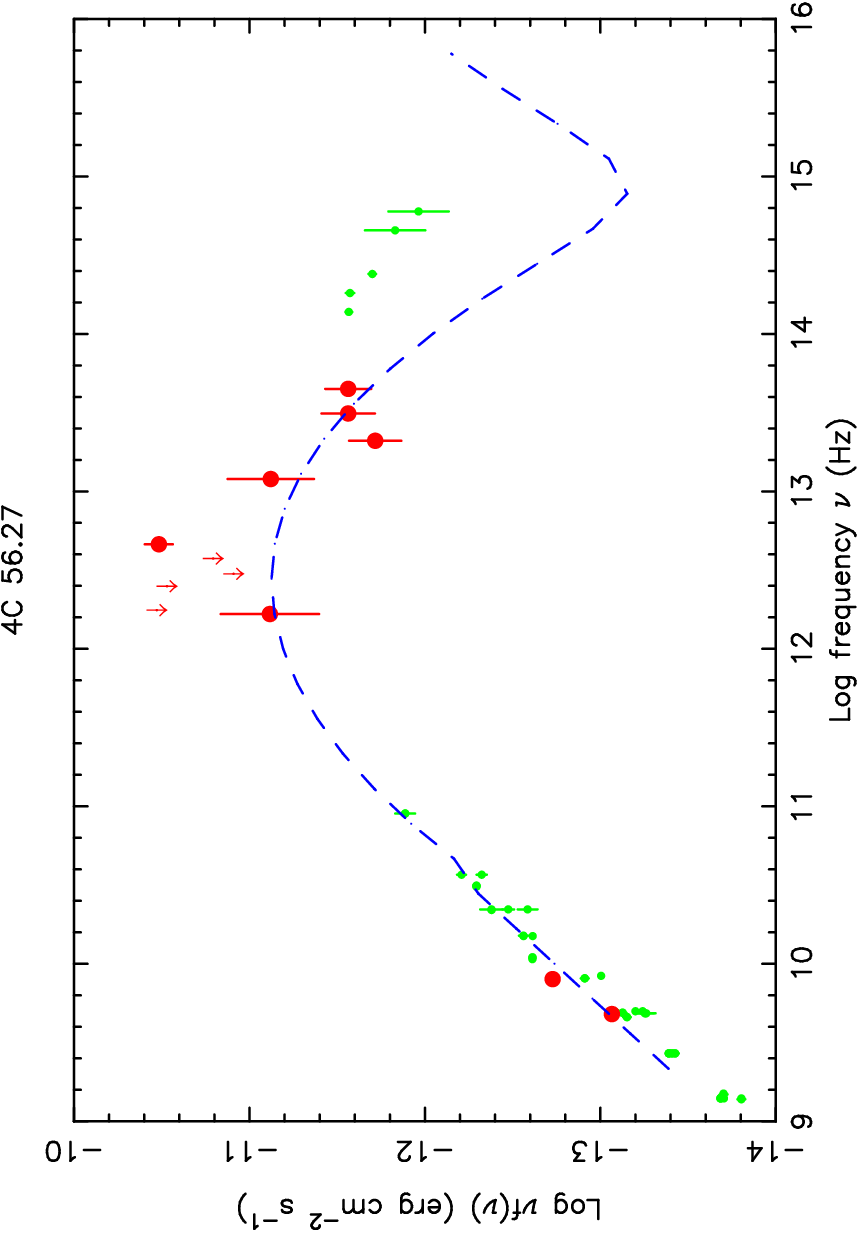} 
\includegraphics{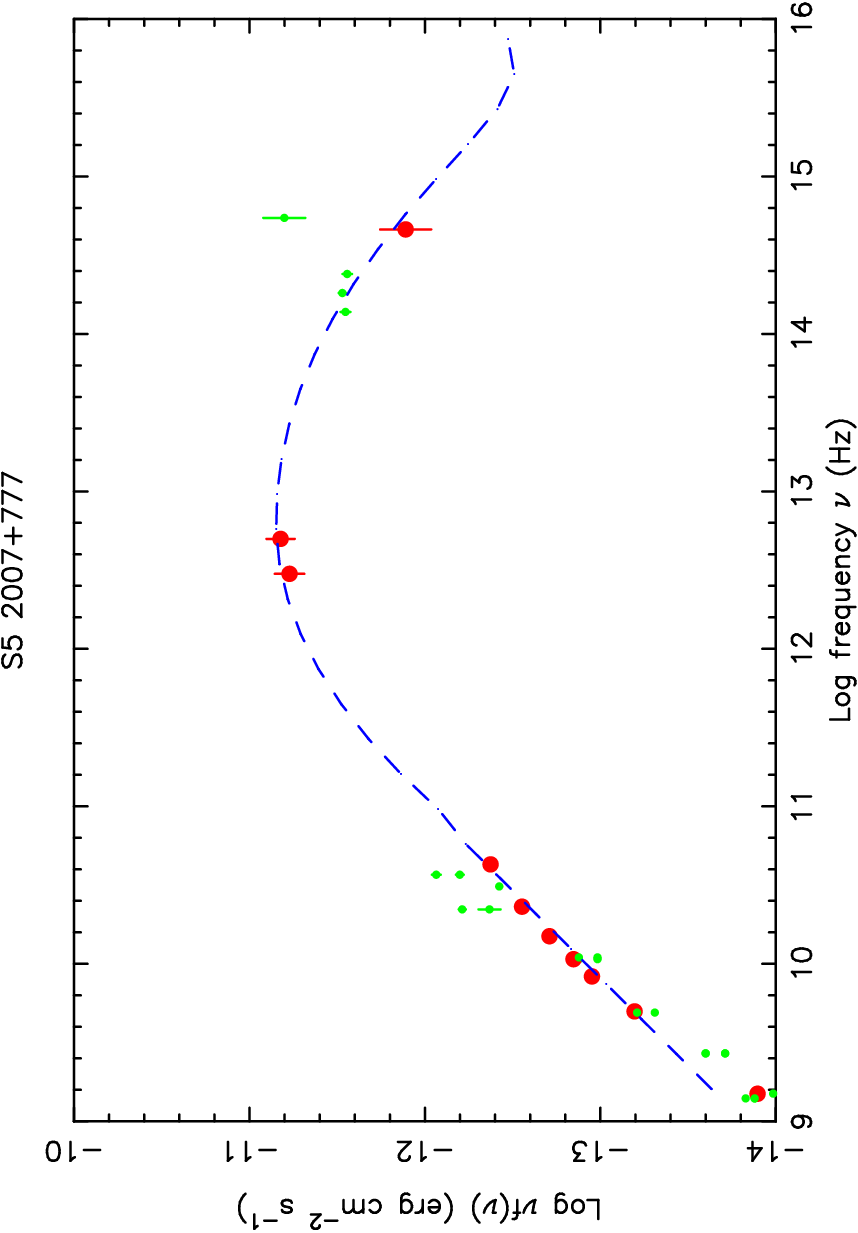} 
\includegraphics{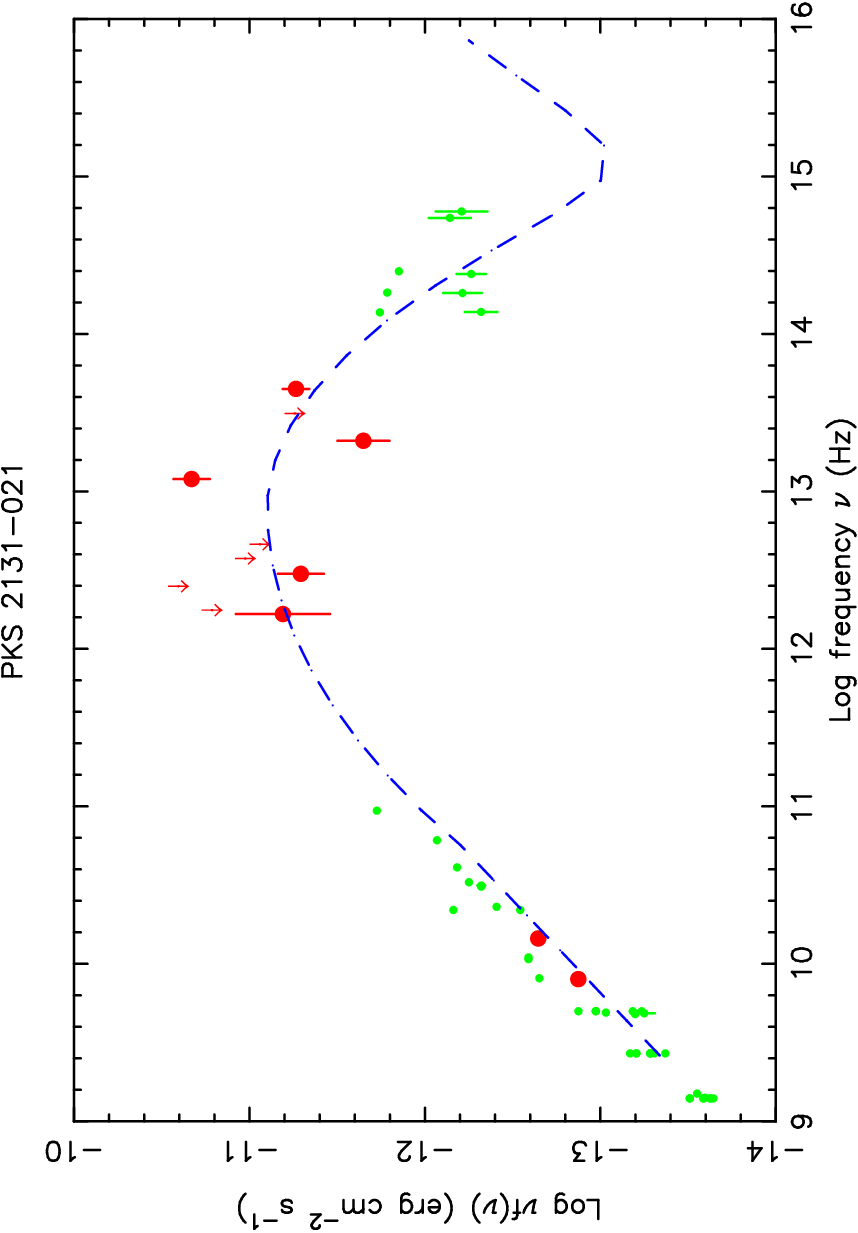} 
\vspace{22.5cm} 
\caption[t]{(continued) The spectral energy distributions for our
sources. Bigger, darker symbols indicate ISO data and the
nearly-simultaneous radio (and optical in the case of PKS 0735+178, OQ 530,
S5 1803+784, and S5 2007+777) data (arrows denote ISO upper limits), while
smaller, lighter symbols represent non-simultaneous data collected from the
literature. The dashed line is our synchrotron model.}\label{seds}
\end{figure*}

\setcounter{figure}{0}
\setcounter{figure}{0}
\begin{figure*}[t]
\centering
\includegraphics{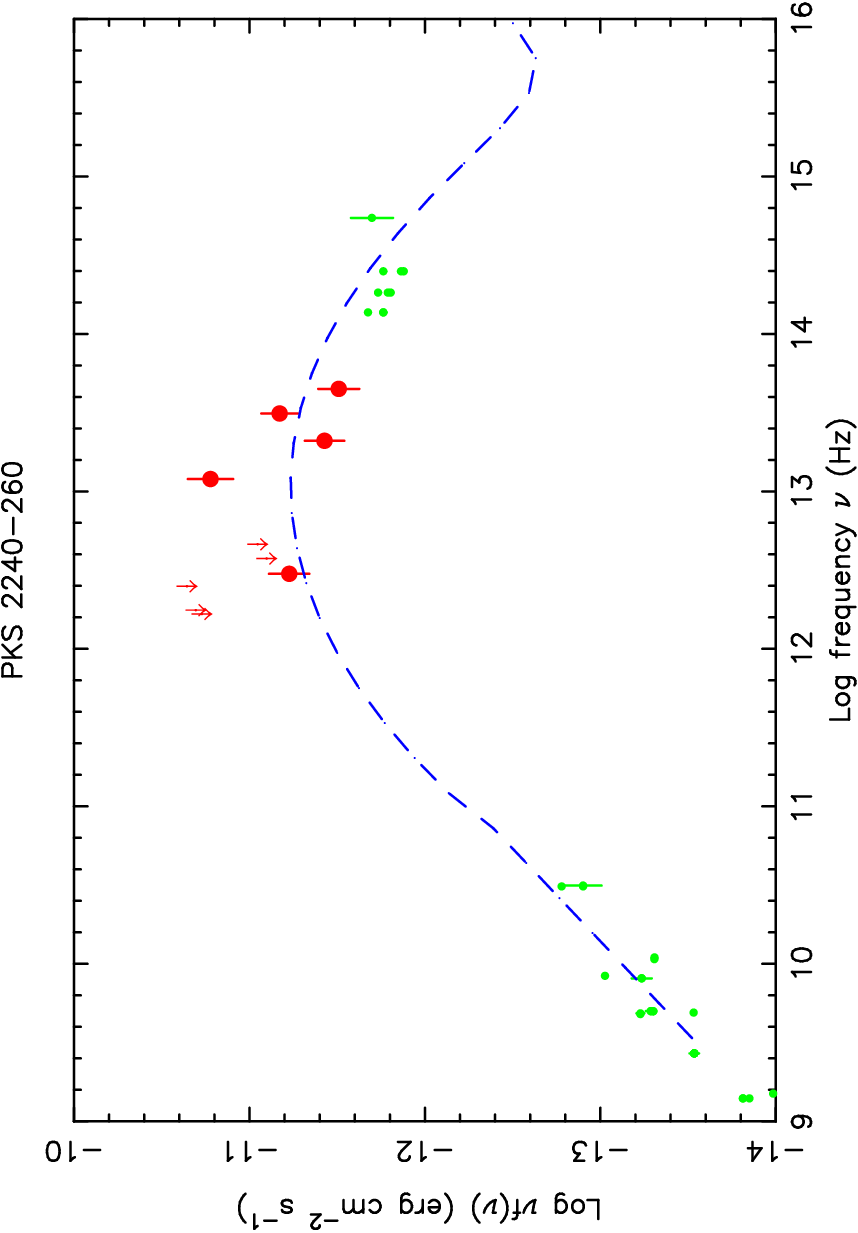} 
\includegraphics{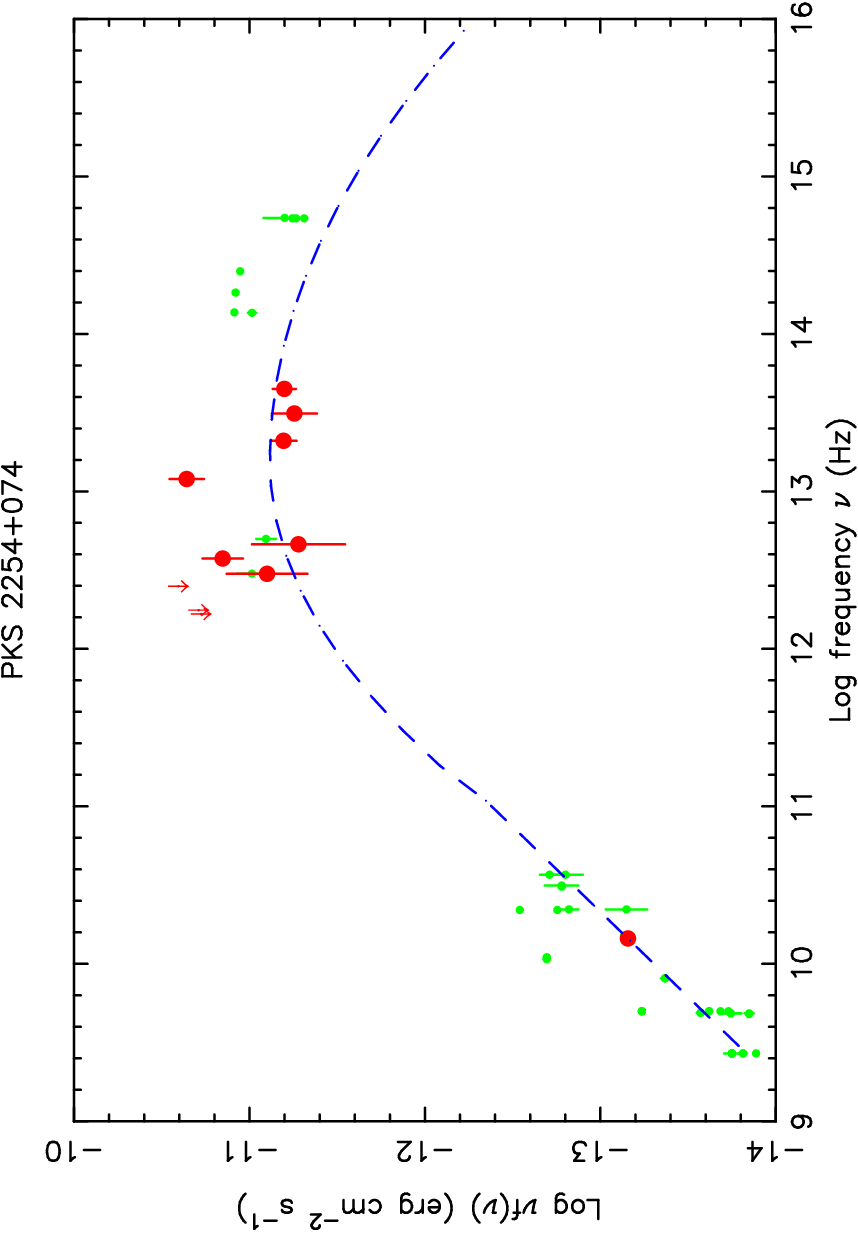} 
\vspace{22.5cm} 
\caption[t]{(continued) The spectral energy distributions for our
sources. Bigger, darker symbols indicate ISO data and the
nearly-simultaneous radio (and optical in the case of PKS 0735+178, OQ 530,
S5 1803+784, and S5 2007+777) data (arrows denote ISO upper limits), while
smaller, lighter symbols represent non-simultaneous data collected from the
literature. The dashed line is our synchrotron model.}\label{seds}
\end{figure*}

\end{document}